\documentclass[twoside,11pt]{article}
\usepackage{jmlr2e}

\usepackage{bm,amsmath,amsfonts}
\usepackage{enumitem}
\usepackage{float}   
\usepackage{rotating}  
\usepackage{pdflscape} 
\usepackage{color}
\usepackage{xcolor}
\usepackage{dsfont}
\usepackage{xurl}
\usepackage{changepage}
\usepackage{multirow,adjustbox}
\usepackage{algorithm}
\usepackage{algorithmic}
\usepackage{booktabs}

\allowdisplaybreaks

\newtheorem{assumption}{Assumption}

\ShortHeadings{DROP: Distributionally Robust Optimization}{Shen, Ji, Li, Yang, and Shi}
\firstpageno{1}

\begin{document}

\title{DROP: Distributionally Robust Optimization for Multi-task Learning in Graphical Models}

\author{\name Canruo Shen \email canruo.shen@ubc.ca \\
       \addr Department of Computer Science, Mathematics, Physics and Statistics \\
       University of British Columbia\\
       Kelowna, V1V 1V7, Canada
       \AND
       \name Xintong Ji \email xintong.ji@connect.polyu.hk \\
       \addr Department of Data Science and Artificial Intelligence\\
       The Hong Kong Polytechnic University\\
       Hung Hom, Kowloon, Hong Kong SAR, China
       \AND
       \name Qiong Li \email qiongli@bnbu.edu.cn \\
       \addr Guangdong Provincial Key Laboratory of Interdisciplinary Research and Application for Data Science \\
       Beijing Normal-Hong Kong Baptist University \\
       Zhuhai, 519087, China 
       \AND
       \name Wenzhi Yang \email wzyang@ahu.edu.cn \\
       \addr School of Big Data and Statistics \\
       Anhui University \\
        Hefei, 230601, China 
       \AND
       \name Xiaoping Shi \email xiaoping.shi@ubc.ca \\
       \addr Department of Computer Science, Mathematics, Physics and Statistics \\
       University of British Columbia\\
       Kelowna, V1V 1V7, Canada }

\editor{My editor}

\maketitle

\begin{abstract}
Gaussian Graphical Models (GGMs) are widely used to infer conditional dependence structures in high-dimensional data. However, standard precision matrix estimators are highly sensitive to data contamination, such as extreme outliers and heavy-tailed noise. In this paper, we propose DROP (\underline{D}istributionally \underline{R}obust \underline{Op}timization), a robust estimation method formulated within a multi-task nodewise regression framework. The proposed estimator enforces structural sparsity while resisting the influence of corrupted observations. Theoretically, we establish error bounds for the DROP estimator under general contamination. Through extensive high-dimensional simulations, we demonstrate that DROP consistently controls the rate of false positive edges and outperforms conventional non-robust estimators when data deviate from standard Gaussian assumptions. Furthermore, in a functional MRI (fMRI) application, DROP maintains a stable graph structure and preserves network modularity even when subjected to severe data perturbations, whereas competing methods yield excessively dense networks. To facilitate reproducible research, the \texttt{DROP} R package will be made publicly available on GitHub.
\end{abstract}

\begin{keywords}
  Gaussian graphical model, Distributionally robust optimization, Nodewise regression, Multi-task learning, Error bounds, fMRI, Precision matrix, High-dimensional
\end{keywords}
·

\section{Introduction} 

Data-driven decision-making frameworks play a central role in a wide range of domains, including machine learning~\citep{sinha2018certifying}, finance~\citep{zhu2009worst}, operations research~\citep{goh2010distributionally}, control~\citep{van2016distributionally}, and statistics~\citep{duchi2021statistics}. In many of these settings, the parameters governing the decision problem are subject to substantial uncertainty arising from limited sample sizes, measurement noise, or predictive inaccuracies~\citep{rahimian2019distributionally}. Such uncertainty requires methods that remain reliable even when the data-generating distribution deviates from nominal assumptions. To address this challenge, Distributionally Robust Optimization (DROP) provides a systematic framework by optimizing performance under the worst-case distribution within a prescribed ambiguity set.

In high-dimensional statistics, Gaussian graphical models (GGMs) are a standard tool for estimating graph structures. However, these models rely heavily on the assumption that the data follow a clean multivariate normal distribution. In practice, real-world data often contain outliers, heavy-tailed errors, or leverage points. When faced with these common anomalies, traditional methods like the Graphical Lasso (GLASSO)~\citep{friedman2008sparse} tend to break down, producing overly dense graphs with many false positive edges. To estimate more reliable graph structures, researchers have applied the DROP framework to GGMs~\citep{cisneros2020distributionally,nguyen2022distributionally}, though these applications have focused on estimating the entire precision matrix globally.

Instead of global estimation, an alternative approach is to recast the precision matrix estimation into a series of node-wise regression tasks~\citep{meinshausen2006high}. Unlike classical approaches that estimate each node's neighborhood independently, aggregating these objectives through scalarization~\citep{boyd2004convex} defines a single joint optimization problem. This formulation explicitly casts GGM estimation as a \emph{multi-task learning} problem~\citep{zhang2025multi}, where each node's regression represents a distinct but interrelated task. However, integrating DROP into this multi-task structure reveals two critical theoretical gaps. First, while Wasserstein DROP~\citep{esfahani2018data} has been successfully used to robustify \emph{single-task} regression models against data perturbations~\citep{blanchet2019robust}, its extension to a dependent multi-task structure remains largely unexplored. Second, existing statistical guarantees for multi-task regression are not formulated under DROP framework. In particular, \citet{lounici2011oracle} assume that the individual tasks are independent. For node-wise GGMs, this assumption fails because the regression tasks are inherently dependent on one another. As a result, current penalties for node-wise GGMs are often chosen out of convenience or empirical observation, rather than being derived from the DROP framework.

\begin{figure}[htbp]
    \centering
    \includegraphics[width=\textwidth]{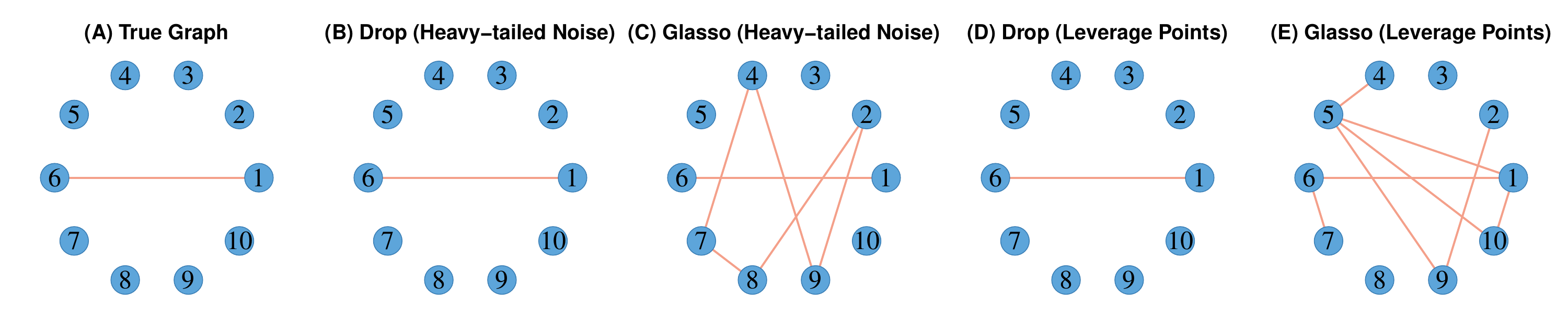}
    \caption{\footnotesize A motivating example illustrating the impact of data contamination on a sparse graph with $p=10$ nodes. 
    (A) The true conditional dependence structure. 
    (B) and (D): Graphs estimated by the proposed DROP method under heavy-tailed noise and leverage points, respectively. 
    (C) and (E): Graphs estimated by the standard GLASSO method under the same contamination scenarios, showing an increase in false positive edges.}
    \label{fig:motivation_example}
\end{figure}

To illustrate the advantage of our approach, Figure~\ref{fig:motivation_example} presents a simulated motivating example on a sparse graph ($p=10, n=5000$); detailed data generation settings are deferred to Section~\ref{sec:simulations}. Panel~(A) displays the true conditional dependence structure. When the observations are perturbed by heavy-tailed noise or leverage points, the standard GLASSO estimator exhibits severe sensitivity, yielding overly dense graphs with a substantial number of false positive edges (Panels~(C) and~(E)). In contrast, our proposed DROP estimator (Panels~B and~D) accommodates these distributional shifts. Through its robust regularization, DROP recovers the underlying sparse structure without introducing additional false positive connections.

This example highlights the critical need for robust estimation procedures that can directly account for distributional uncertainty and maintain structural integrity under data contamination. To address the aforementioned gaps, our research investigates three central questions: 
(i) How can Wasserstein DROP principles and the multi-task nature of node-wise GGM estimation be combined to create a generalized, robust estimator? 
(ii) What are the fundamental statistical properties and convergence rates of parameter estimation when extending the DROP framework from single-task regression to multi-task learning setups? 
(iii) Does this theoretical framework translate into practical computational scalability and better empirical recovery in high-dimensional settings?

In this paper, we answer these questions and make the following primary contributions:
\begin{itemize}
    \item \textbf{The DROP Estimator:} We propose the DROP method, which leverages scalarization to aggregate node-wise regressions into a joint \emph{multi-task learning} framework. By applying Wasserstein DROP to this unified objective, the adaptive robust regularizer is systematically derived rather than chosen empirically.
    \item \textbf{Theoretical Foundation:} We establish the framework for deriving specific penalty functions that naturally arise from the DROP formulation of regression loss functions. We prove error bounds for these estimators in both single-task and multi-task learning setups with high probability.
    \item \textbf{Empirical Performance and R Package:} Through extensive simulations and an application to real-world functional MRI (fMRI) data, we demonstrate that DROP maintains robust graph recovery and computational efficiency under severe data contamination. To facilitate reproducibility, the proposed method is implemented in the R package \texttt{DROP}, which will be made publicly available on GitHub.
\end{itemize}

The rest of the paper is organized as follows: Section~\ref{sec:background} provides the background on Wasserstein DROP and a review of existing GGM estimators. Section~\ref{sec:DROP} introduces the proposed DROP methodology for GGMs, detailing the derivation of the robust penalty. Section~\ref{sec:methodology} establishes the error bounds for parameter estimators under the DROP framework in both single-task and multi-task learning setups. Section~\ref{sec:algorithm} details the accompanying optimization algorithm. Section~\ref{sec:simulations} presents our simulation studies, followed by a real-world fMRI data application in Section~\ref{sec:real_data}. Finally, we conclude with a discussion in Section~\ref{sec:discussion}. The detailed proofs of the theorems, technical lemmas, and additional simulation results are provided in the Appendix.

\section{Background and Related Work}\label{sec:background}

\subsection{Gaussian Graphical Model Estimation}
Gaussian Graphical Models (GGMs) offer a powerful way to visualize and model conditional dependence relationships among a set of $p$ variables. We assume these variables, denoted as the random vector $X = (X_1, \dots, X_p)^\top$, follow a $p$-dimensional multivariate Gaussian distribution, $\mathcal{N}(\mathbf{0}, \Sigma)$, where $\Sigma \in \mathbb{R}^{p \times p}$ is the covariance matrix. Let the $n \times p$-dimensional data matrix $\mathbf{X}$ contain $n$ independent observations of $X$, so that the columns $\mathbf{X}_i$ correspond for all $i \in \{1, \dots, p\}$ to the vector of $n$ independent observations of $X_i$. We assume the data matrix $\mathbf{X}$ has been centred so that each column has zero mean. The empirical covariance matrix corresponding to $\Sigma$ is $S := \frac{1}{n} \mathbf{X}^\top \mathbf{X}$. 

In a GGM, the relationships between variables are represented by an undirected graph $G = (V, E)$, where the nodes $V = \{1, \dots, p\}$ correspond to the variables. The absence of an edge between nodes $i$ and $j$ in the edge set $E$ signifies the conditional independence of variables $X_i$ and $X_j$ given all other variables. The core theoretical relationship is formalized through the precision matrix, also known as the inverse covariance matrix, $K := \Sigma^{-1} \in \mathbb{R}^{p \times p}$. Specifically, for variables $X$ following a $\mathcal{N}(\mathbf{0}, \Sigma)$ distribution, $K_{ij} = 0$ if and only if $X_i$ and $X_j$ ($i \neq j$) are conditionally independent given the remaining variables. Consequently, estimating the sparse graph structure is equivalent to estimating a sparse precision matrix $K$. This task becomes challenging in high-dimensional settings, where the number of variables $p$ may exceed the sample size $n$. In such cases, the sample covariance matrix $S$ is singular and cannot be directly inverted. Regularization techniques are thus essential to obtain stable and sparse estimates of $K$.

Our discussion focuses primarily on frequentist methods for GGMs. From a methodological viewpoint, existing frequentist approaches to sparse Gaussian graphical model estimation can be grouped into two complementary families~\citep{fan2016overview}: global penalized likelihood estimators and column-wise precision matrix estimators.

\subsubsection{Global penalized likelihood estimators}
A major line of frequentist methodology for GGMs is based on penalized likelihood estimation. These approaches estimate the entire precision matrix $K$ jointly by optimizing a global objective of the form
\begin{equation}
    \hat{K}
    =\arg\min_{K\succ 0}\Big\{
    Q(S,K)+\text{Pen}_\lambda(K)
    \Big\},
\end{equation}
where $S$ is the empirical covariance matrix, $Q(S,K)= -\log\det(K)+\operatorname{tr}(SK)$ denotes the Gaussian negative log-likelihood, and $\text{Pen}_\lambda(K)$ is a matrix-valued regularizer promoting sparsity.
The Graphical Lasso (GLASSO)~\citep{banerjee2008model,yuan2007model,friedman2008sparse} employs the elementwise $\ell_1$-penalty,
$\text{Pen}^{\text{GLASSO}}_\lambda(K)=\lambda\|K\|_1$,
yielding the convex program
\begin{equation}
 \hat{K}^{\text{GLASSO}}
=\arg\min_{K\succ 0}
\left\{-\log\det(K)+\operatorname{tr}(SK)+\lambda\|K\|_1\right\}.   
\end{equation}
This joint estimator differs from column-wise methods by enforcing global symmetry and positive definiteness throughout the optimization. Efficient block coordinate descent algorithms~\citep{banerjee2008model} have made GLASSO a standard benchmark for sparse precision matrix estimation.

To mitigate estimation bias introduced by the $\ell_1$-penalty, subsequent work explored a range of non-convex regularizers, including early contributions by~\citet{lam2009sparsistency} and~\citet{johnson2012high}, as well as folded-concave penalties such as smoothly clipped absolute deviation (SCAD)~\citep{fan2001variable} and minimax concave penalty (MCP)~\citep{zhang2010nearly}. In our setting, these non-convex penalties are applied elementwise to the off-diagonal entries $K_{ij}$ with $i\neq j$ of the precision matrix. For SCAD, with regularization parameter $\lambda>0$ and shape parameter $a>2$ (often $a=3.7$), the contribution of an individual entry is
\begin{equation}
    \text{pen}^{\text{SCAD}}_\lambda(|K_{ij}|)=
    \begin{cases}
    \lambda |K_{ij}|, & 0\le |K_{ij}|\le \lambda,\\[4pt]
    \displaystyle\frac{-|K_{ij}|^2+2a\lambda |K_{ij}|-\lambda^2}{2(a-1)}, & \lambda< |K_{ij}|\le a\lambda,\\[8pt]
    \displaystyle\frac{(a+1)\lambda^2}{2}, & |K_{ij}|>a\lambda,
    \end{cases}
\end{equation}
so that the SCAD penalty functional on the precision matrix is
\[
\text{Pen}^{\text{SCAD}}_\lambda(K)
=\sum_{i\neq j} \text{pen}^{\text{SCAD}}_\lambda(|K_{ij}|).
\]
For MCP, with shape parameter $a>1$, we similarly define the penalty
\begin{equation}
    \text{pen}^{\text{MCP}}_\lambda(|K_{ij}|)=
    \begin{cases}
    \lambda |K_{ij}|-\dfrac{|K_{ij}|^2}{2a}, & 0\le |K_{ij}|\le a\lambda,\\[8pt]
    \displaystyle\dfrac{a\lambda^2}{2}, & |K_{ij}|>a\lambda,
    \end{cases}    
\end{equation}
and the corresponding matrix-valued penalty
\[
    \text{Pen}^{\text{MCP}}_\lambda(K)
    =\sum_{i\neq j} \text{pen}^{\text{MCP}}_\lambda(|K_{ij}|).
\]
These closed-form piecewise representations are obtained by integrating the standard SCAD and MCP derivatives and are given by~\citet{breheny2011coordinate}. 
This unified penalized-likelihood framework supports a wide class of regularizers, enabling additional structure, such as robustness considerations, to be incorporated through the design of $\text{Pen}_\lambda(\cdot)$. 

\subsubsection{Column-wise precision matrix estimators}
Beyond global likelihood methods, an alternative line of work constructs the precision matrix column by column. Column-wise methods can be broadly grouped into two families: (i) graphical Dantzig selector~\citep{candes2007dantzig} type estimators, and (ii) Lasso type regression approaches. We first discuss the former category.~\citet{cai2011constrained} proposed the CLIME estimator, which directly estimates the $i$-th column of precision matrix $K$, denoted by $K_i$, via
\begin{equation}
    \hat{K}^{\text{CLIME}}_{i}
    =\arg\min_{K_i} \|K_i\|_1
    \quad \text{subject to} \quad
    \|S K_i - e_i\|_\infty \le \zeta_i,
    \qquad i=1,\ldots,p,
\end{equation}
where $e_i$ is the $i$-th canonical vector and $\zeta_i$ is a tuning parameter.~\citet{cai2011constrained} show that this convex program can be reformulated as a linear optimization problem.
After obtaining $\hat K$, a secondary thresholding step is typically applied to recover the edge set of the underlying graph. In subsequent work,~\citet{liu2015fast} proposed the SCIO estimator, which also estimates the $i$-th column of $K$ independently.
Specifically, SCIO solves the convex problem
\begin{equation}
    \hat{K}^{\text{SCIO}}_{i}
    =\arg\min_{K_i}\left\{
    \tfrac{1}{2} K_i^\top S K_i
    - e_i^\top K_i
    + \lambda_i \|K_i\|_1
    \right\},
    \qquad i=1,\ldots,p,
\end{equation}
where $\lambda_i$ is a regularization parameter controlling sparsity. 
This quadratic objective provides a smooth surrogate for the population equation $\Sigma K_i = e_i$, enabling efficient column-wise estimation.

The second major family of column-wise estimators is based on Lasso type regressions, commonly referred to as node-wise regression approaches. These methods adopt a node-wise regression perspective, reconstructing the precision matrix $K$ by solving $p$ separate sparse regression problems. Because each regression corresponds to a distinct estimation objective, this decomposition naturally induces a multi-task optimization structure. This idea can be traced back to the Neighborhood Selection (NS) framework proposed by~\citet{meinshausen2006high}. Instead of estimating the entire precision matrix $K$ jointly, this method decomposes the challenge into $p$ independent linear regression problems. For each node $i$, the $n$ observations of the variable $X_i$ form the $i$-th column of $\mathbf{X}$, denoted $\mathbf{X}_i \in \mathbb{R}^n$. The goal is to predict $\mathbf{X}_i$ using the remaining columns $\mathbf{X}_{-i} \in \mathbb{R}^{n \times (p-1)}$ (the matrix $\mathbf{X}$ with the $i$-th column removed). Based on the conditional properties of the Gaussian distribution, the non-zero coefficients in the theoretical regression of $X_i$ onto $X_{-i}$ correspond to the non-zero off-diagonal elements in the $i$-th row (or column) of the precision matrix $K$.

A general formulation for this class of node-wise estimators takes the form
\begin{equation}
   \hat{\boldsymbol{\beta}}^{(i)}=\arg \min _{\boldsymbol{\beta}^{(i)} \in \mathbb{R}^{p-1}}\left\{L\left(\mathbf{X}_i, \mathbf{X}_{-i}, \boldsymbol{\beta}^{(i)}\right)+\text{Pen}_\lambda\left( \boldsymbol{\beta}^{(i)} \right)\right\}, 
\end{equation}
where $L(\cdot)$ denotes a chosen loss function and $\text{Pen}_\lambda(\cdot)$ is a regularization term parameterized by $\lambda$. The graph structure is then inferred by combining the estimated neighborhoods, often by declaring an edge between $i$ and $j$ if both $\hat{\beta}_j^{(i)}$ and $\hat{\beta}_i^{(j)}$ are non-zero. The theoretical justification relies on the identity $\beta_j^{(i)} = -K_{ij}/K_{ii}$ relating the coefficients of the theoretical regression of $X_i$ onto $X_{-i}$ to the elements of the precision matrix $K$. Our proposed method later builds upon this idea, defining a related vector $\boldsymbol{\beta}^{(i)}(K)$ such that $\beta^{(i)}_j(K) = -K_{ij}/K_{ii}$ for $j\neq i$.

Building on this node-wise regression framework, a substantial body of work has explored different specifications of the loss function $L(\cdot)$ and penalty $\text{Pen}_\lambda(\cdot)$ for reconstructing sparse precision matrices. Neighborhood Selection (NS)~\citep{meinshausen2006high} employs the classical least-squares loss to model the conditional mean of each variable and introduces sparsity through an $\ell_1$ penalty:
\begin{equation}
    L^{\text{NS}}(\mathbf{X}_i,\mathbf{X}_{-i},\boldsymbol{\beta}^{(i)}(K))
    = \frac{1}{n}\|\mathbf{X}_i - \mathbf{X}_{-i}\boldsymbol{\beta}^{(i)}(K)\|_2^2,
    \qquad
    \text{Pen}^{\text{NS}}_\lambda(\boldsymbol{\beta}^{(i)}(K)) = \lambda\|\boldsymbol{\beta}^{(i)}(K)\|_1.
\end{equation}
Scaled-Lasso~\citep{sun2013sparse} modifies the least-squares loss by introducing an additional scale parameter $\sigma_i=\left(K_{i i}\right)^{-1 / 2}$, which is jointly optimized with the regression coefficients:
\begin{equation}
\begin{aligned}
    L^{\text{SCALED}}(\mathbf{X}_i, \mathbf{X}_{-i}, \boldsymbol{\beta}^{(i)}(K), \sigma_i) &= \frac{1}{2n\sigma_i}\|\mathbf{X}_i - \mathbf{X}_{-i}\boldsymbol{\beta}^{(i)}(K)\|_2^2 + \frac{\sigma_i}{2}, \\
    \text{Pen}^{\text{SCALED}}_\lambda(\boldsymbol{\beta}^{(i)}(K)) &= \lambda\|\boldsymbol{\beta}^{(i)}(K)\|_1.
\end{aligned}
\end{equation}
Tuning-Insensitive Graph Estimation and Regression (TIGER)~\citep{liu2017tiger} further modifies the objective by replacing the quadratic loss with the Square-Root Lasso:
\begin{equation}
\begin{aligned}
    L^{\text{TIGER}}(\mathbf{X}_i, \mathbf{X}_{-i}, \boldsymbol{\beta}^{(i)}(K)) &= \sqrt{\frac{1}{n}\|\mathbf{X}_i - \mathbf{X}_{-i}\boldsymbol{\beta}^{(i)}(K)\|_2^2}, \\
    \text{Pen}^{\text{TIGER}}_\lambda(\boldsymbol{\beta}^{(i)}(K)) &= \lambda\|\boldsymbol{\beta}^{(i)}(K)\|_1.
\end{aligned}
\end{equation}

Both global and column-wise estimators fundamentally assume that the underlying data follow a multivariate Gaussian distribution. However, real-world datasets frequently contain heavy-tailed noise or leverage points. Because standard methods are highly sensitive to such deviations, their estimation performance often deteriorates significantly, highlighting the need for robust alternatives.

\subsection{Robust and Nonparanormal Estimators}
To address the vulnerability of standard GGMs to distributional irregularities, the Nonparanormal (NPN) framework~\citep{liu2009nonparanormal,liu2012high,xue2012regularized} was proposed to relax the strict normality assumption using semi-parametric Gaussian copula models. The NPN model assumes that while the random vector $X = (X_1, \dots, X_p)^\top$ may not be Gaussian, there exists a set of monotone transformation functions $\{f_i\}_{i=1}^p$ such that the transformed vector $f(X) = (f_1(X_1), \dots, f_p(X_p))^\top$ follows a multivariate normal distribution $\mathcal{N}(\mathbf{0}, \Sigma)$. The conditional independence structure is then encoded in the precision matrix $K := \Sigma^{-1}$ of this latent Gaussian distribution.

Methodologically, estimation within the NPN framework generally follows two main strategies. The first strategy, originally introduced by~\citet{liu2009nonparanormal}, involves an element-wise transformation that explicitly estimates the underlying transformation functions $f_i$. A standard approach is to use the empirical cumulative distribution function (ECDF) combined with a normal quantile transformation:
\begin{equation}
    \hat{f}_i(x) = \Phi^{-1}(\hat{F}_i(x)),
\end{equation}
where $\Phi^{-1}$ is the inverse standard normal CDF and $\hat{F}_i$ is the truncated ECDF of the $i$-th variable. The transformed data matrix is then denoted by $\tilde{\mathbf{X}}$, whose columns are obtained by applying the estimated transformations element-wise, such that $\tilde{\mathbf{X}}_i = \hat{f}_i(\mathbf{X}_i)$ for each $i=1,\dots,p$.

The second strategy, explicitly referred to as the "Nonparanormal Skeptic"~\citep{liu2012high}, bypasses the explicit estimation of marginal transformations and instead estimates the latent correlation matrix directly using rank-based statistics. This approach exploits the functional relationship between rank correlation coefficients and the latent Pearson correlation. For Kendall's $\tau$, the element-wise robust correlation estimator, denoted as $\hat{R}^{\tau}$, is given by:
\begin{equation}
    \hat{R}_{ij}^{\tau} = \sin\left(\frac{\pi}{2} \hat{\tau}_{ij}\right),
\end{equation}
where $\hat{\tau}_{ij}$ is the empirical Kendall rank correlation coefficient between variables $X_i$ and $X_j$. Similarly, using Spearman's rank correlation~\citep{xue2012regularized}, denoted here by $\eta$, the robust correlation matrix $\hat{R}^{\eta}$ can be estimated via:
\begin{equation}
    \hat{R}_{ij}^{\eta} = 2\sin\left(\frac{\pi}{6} \hat{\eta}_{ij}\right),
\end{equation}
where $\hat{\eta}_{ij}$ denotes the empirical Spearman's rank correlation. Theoretical studies~\citep{liu2012high,xue2012regularized} demonstrate that substituting the standard sample covariance matrix $S$ with these rank-based estimators ($\hat{R}^{\tau}$ or $\hat{R}^{\eta}$) in algorithms like GLASSO yields consistent graph recovery, even when the underlying data distribution is heavy-tailed.

Although nonparanormal methods help relax the Gaussian assumption on marginal distributions, they are essentially element-wise preprocessing steps. Consequently, they may not fully resolve the estimation instability caused by severe outliers or complex data contamination. This limitation suggests that a robust optimization framework is needed during the model estimation stage to handle the underlying distributional uncertainty.

\subsection{Distributionally Robust Optimization Formulation}
A general DROP problem provides a principled way to handle distributional uncertainty~\citep{delage2010distributionally}. It is formulated as:
\begin{equation}
\hat{\theta}_n = \min_{\theta \in {\Theta}} \sup_{\mathbb{P} \in \mathcal{D}} \mathbb{E}_{\mathbb{P}}[l(z;\theta)], 
\label{eq:dro_general}
\end{equation}
where $\theta \in {\Theta}$ is the model parameter, $z$ is a random variable governed by a probability distribution $\mathbb{P}$ from within an ambiguity set $\mathcal{D}$, and $l(z;\theta)$ is the loss function. The objective is to minimize the expected loss under the worst-case distribution $\mathbb{P} \in \mathcal{D}$. This formulation appears in foundational works such as~\citet{rahimian2019distributionally,kuhn2025distributionally}. Compared to classical stochastic optimization, DROP explicitly protects against model misspecification and improves generalization in the presence of data variability~\citep{wiesemann2014distributionally,goh2010distributionally}.

In Wasserstein DROP~\citep{esfahani2018data}, the ambiguity set $\mathcal{D}$ in \eqref{eq:dro_general} is defined as a ball centered around a nominal distribution, typically the empirical distribution $\mathbb{P}_n$ derived from the observed data $z_1, \dots, z_n$. The size of this ball is measured by the $q$-Wasserstein distance, $W_q$ (for $q \ge 1$). The $q$-Wasserstein distance between distributions $\mathbb{P}$ and $ \mathbb{P}_n$ is defined as:
\begin{equation}
W_q(\mathbb{P}, \mathbb{P}_n) = \left( \inf_{\pi \in \Pi(\mathbb{P}, \mathbb{P}_n)} \mathbb{E}_{(z, \tilde{z}) \sim \pi} \left[ d(z, \tilde{z})^q \right] \right)^{1/q},
\label{eq:wdistance}
\end{equation}
where $\Pi(\mathbb{P}, \mathbb{P}_n)$ is the set of all couplings (joint distributions) with marginals $\mathbb{P}$ and $\mathbb{P}_n$, and $d(z, \tilde{z})$ is a chosen ground metric. This distance quantifies the minimal expected cost of transporting mass to transform one distribution into the other \citep{panaretos2019statistical}. The core Wasserstein DROP problem involves minimizing the worst-case expected loss over this ambiguity set~\citep{kuhn2025distributionally,gao2024wasserstein}:
\begin{equation}
\min_{\theta \in {\Theta}} \sup_{\mathbb{P} : W_q(\mathbb{P}, \mathbb{P}_n) \leq \rho} \mathbb{E}_{z \sim \mathbb{P}}[l(z;\theta)]
\label{wro_empirical}
\end{equation}
Here, $l( \cdot)$ represents the loss function and $\rho > 0$ is the radius dictating the level of robustness. Because the true data-generating distribution is unknown and only empirical samples are observed, Problem~\eqref{wro_empirical} adopts a minimax formulation centered at $\mathbb{P}_n$.

A foundational property of Wasserstein DROP is its deep connection to regularized optimization. \citet{gao2024wasserstein} show that, under certain conditions, the worst-case expectation~\eqref{wro_empirical} can be approximated via a first-order expansion of the loss function around $\mathbb{P}_n$. This leads to a regularized empirical risk minimization problem:
\begin{equation}
\hat{\theta}_n \approx  \arg\min_{\theta \in {\Theta}} \left\{ \mathbb{E}_{z \sim \mathbb{P}_n}[l(z;\theta)] + \rho_n \cdot V_{\mathbb{P}_n, q^*}(l(z;\theta)) \right\},
\end{equation}
where $\rho_n$ is a scaled radius, $q^*$ is the Hölder conjugate of $q$ ($1/q + 1/q^* = 1$), and $V_{\mathbb{P}_n, q^*}(l(z;\theta))$ is the variation of the loss defined by~\citet{gao2024wasserstein}.

This duality becomes particularly powerful when applied to specific predictive tasks, such as linear regression, which forms the foundational building block of our node-wise GGM estimator DROP. Specifically,~\citet{blanchet2019robust} established that under a Wasserstein DROP framework, minimizing the worst-case expected square loss is analytically equivalent to a regularized regression problem. They consider a discrepancy measure $\mathcal{D}_c$, where $\mathcal{D}_c^{\,1/q}$ corresponds to the Wasserstein distance of order $q$ ($W_q$). In the specific case of the $2$-Wasserstein distance ($q=2$), when the transportation cost restricts adversarial perturbations strictly to the predictor variables via an $\ell_{\kappa^*}$-norm, the DROP problem exactly reduces to the Square-root Lasso formulation. Let $Y \in \mathbb{R}$ denote a generic response variable and $X^{(g)} \in \mathbb{R}^d$ denote a generic predictor vector. The equivalence is given by:
\begin{equation}
    \inf_{\boldsymbol{\beta} \in \mathbb{R}^d} \sup_{\mathbb{P}: \mathcal{D}_c(\mathbb{P}, \mathbb{P}_n) \le \delta} \mathbb{E}_{\mathbb{P}}[(Y - \boldsymbol{\beta}^\top X^{(g)})^2] = \inf_{\boldsymbol{\beta} \in \mathbb{R}^d} \left\{ \sqrt{\mathrm{MSE}_n(\boldsymbol{\beta})} + \sqrt{\delta} \|\boldsymbol{\beta}\|_{\kappa^*} \right\}^2,
\label{eq:dro_equivalence}
\end{equation}
where $\mathrm{MSE}_n(\boldsymbol{\beta}) = \frac{1}{n} \sum_{k=1}^n (y_k - \boldsymbol{\beta}^\top x_k^{(g)})^2$ is the empirical mean squared error computed over $n$ independent observations $\{(y_k, x_k^{(g)})\}_{k=1}^n$. Here, $\delta$ dictates the radius of the uncertainty set, and $\|\boldsymbol{\beta}\|_{\kappa^*} $ denotes the $\ell_{\kappa^*} \text {-norm}$. The regularization parameter $\lambda$ in the corresponding penalized regression is thus explicitly given by $\sqrt{\delta}$, providing a direct mapping between the level of distributional robustness and the degree of model sparsity.

This link is central to precision matrix estimation. It shows that the robust penalty is not an arbitrary choice, but a direct mathematical result of protecting the model against distributional shifts. This framework allows us to develop regularizers that naturally suit the multi-task structure of GGM estimation.

\subsection{Distributionally Robust Estimators for GGMs}
Building on this equivalence, recent studies have applied DROP to address model uncertainty in GGMs. Unlike standard penalized likelihood or column-wise methods, which optimize an objective based solely on the empirical distribution $\mathbb{P}_n$, DROP approaches find an estimator that minimizes the loss under the worst-case distribution within a neighborhood of $\mathbb{P}_n$. This approach explicitly accounts for finite-sample distributional shifts and data contamination~\citep{fan2016overview,rahimian2019distributionally}.

For global precision matrix estimation,~\citet{nguyen2022distributionally} proposed the Wasserstein Shrinkage Estimator, establishing theoretical guarantees under measure concentration conditions. In this setting, the radius of the Wasserstein ambiguity set determines the regularization parameter. Unlike standard empirical $\ell_1$ penalties, the regularization is directly derived from the geometry of the uncertainty set, offering inherent protection against data perturbations at the inference stage.

This perspective aligns well with the nonparanormal methods discussed in Section 2.2. While NPN methods focus on transforming marginal distributions, DROP accommodates distributional ambiguity by modifying the optimization objective itself~\citep{fan2016overview}. By determining the penalty structure through the uncertainty set, this formulation connects data robustness with regularization~\citep{blanchet2025distributionally,cisneros2020distributionally}.

However, while DROP has been applied to global penalized likelihood estimation, its integration into the node-wise regression framework remains largely unexplored. Specifically, adapting the DROP penalty to the multi-task structure of node-wise GGM estimation represents an important gap in the literature. In the following section, we address this gap by proposing the DROP methodology and establishing its theoretical properties.

\section{The DROP Methodology}
\label{sec:DROP}

Building on the distributionally robust optimization concepts reviewed in Section~2, we now formally derive the DROP methodology. In this section, we construct the robust joint estimation framework and detail its computational algorithm. The error bound theoretical guarantees underpinning this approach will be established subsequently in Section~\ref{sec:methodology}.

\subsection{Model Overview}

Building on the node-wise regression perspective reviewed in Section~1, the proposed DROP methodology aggregates $p$ neighborhood selection~\citep{meinshausen2006high} objectives through scalarization~\citep{boyd2004convex}, and incorporates robustness against distributional perturbations via Wasserstein distributionally robust optimization.

Unlike classical neighborhood selection methods that estimate each node's neighborhood independently, our formulation defines a single joint optimization problem that balances empirical fit and robustness across all nodes simultaneously. This naturally casts GGM estimation as a multi-task learning problem, where each node-wise regression represents a distinct task with its own sensitivity to distributional uncertainty. The explicit form of the objective function is derived in the following subsection by applying the scalarization principle and the Wasserstein DROP framework of~\citet{blanchet2019robust} to the node-wise loss functions.


\subsection{Distributionally Robust Scalarized Multi-Task GGM}
We begin with the general distributionally robust optimization formulation introduced in Section~2. Let $(X^{(g)}, Y)$ denote a generic observation composed of predictor variables $X^{(g)}$ and a response $Y$, and let $l(X^{(g)}, Y; \boldsymbol{\beta})$ be the associated loss function. We consider the inner maximization problem
\begin{equation}
\sup_{\mathbb{P}:\, \mathcal{D}_c(\mathbb{P}, \mathbb{P}_n) \le \delta}
\mathbb{E}_{\mathbb{P}}\big[l(X^{(g)}, Y; \boldsymbol{\beta})\big],
\end{equation}
where $\mathbb{P}_n$ denotes the empirical distribution and $\mathcal{D}_c$ is an optimal transport discrepancy induced by a lower semicontinuous cost function $c$. 

Following the dual reformulation of Wasserstein distributionally robust optimization \citep{blanchet2019robust}, we introduce a nonnegative Lagrange multiplier $\gamma \ge 0$ associated with the discrepancy constraint, yielding the Lagrangian
\begin{equation}
\mathcal{L}(\mathbb{P},\gamma)
=
\mathbb{E}_{\mathbb{P}}\big[l(X^{(g)}, Y; \boldsymbol{\beta})\big]
-
\gamma\big(\mathcal{D}_c(\mathbb{P}, \mathbb{P}_n) - \delta\big).
\end{equation}
Under the regularity conditions stated in Section~2, strong duality holds, allowing the interchange of the supremum over probability measures $\mathbb{P}$ and the infimum over $\gamma$.

To proceed, we make the structure of the optimal transport discrepancy explicit. Let $P_1$ and $P_2$ be probability measures supported on $\mathbb{R}^{d}\times\mathbb{R}$. Given a lower semicontinuous cost function
$c : (\mathbb{R}^{d}\times\mathbb{R}) \times (\mathbb{R}^{d}\times\mathbb{R})
\to [0,\infty)$, define
\begin{equation}
\mathcal{D}_c(P_1, P_2)
=
\inf_{\pi \in \Pi(P_1, P_2)}
\mathbb{E}_{\pi}
\big[
c\big((U,V),(U',V')\big)
\big],
\end{equation}
where $\Pi(P_1, P_2)$ denotes the set of all couplings of $P_1$ and $P_2$. Generally, choosing $c((u,v), (u',v')) = \| (u,v) - (u',v') \|^q$ induces the $q$-Wasserstein distance. For the distributionally robust framework developed in this work, we specifically require the $2$-Wasserstein distance. Therefore, we restrict the transportation cost to the squared $\ell_\kappa$-distance (for $\kappa >1$). That is, for any two points $(u,v)$ and $(u',v')$ in $\mathbb{R}^{d}\times\mathbb{R}$, we set
\begin{equation}
c\big((u,v), (u',v')\big) = \big\| (u,v) - (u',v') \big\|_\kappa^2.
\end{equation}

This representation implies that the effect of the distributional constraint $\mathcal{D}_c(\mathbb{P}, \mathbb{P}_n)\le \delta$ can be analyzed locally around each empirical observation. In particular, the contribution of a single data point $(x_k^{(g)}, y_k)$ is captured by the function $\phi_\gamma(\cdot)$
\begin{equation}
\phi_\gamma(x_k^{(g)}, y_k; \boldsymbol{\beta})
:=
\sup_{(u,v)\in\mathbb{R}^{d}\times\mathbb{R}}
\left\{
l(u,v;\boldsymbol{\beta})
-
\gamma\,
c\big((u,v),(x_k^{(g)}, y_k)\big)
\right\}.
\label{phi_gamma}
\end{equation}
As a result, the original infinite-dimensional optimization problem over probability measures reduces to the finite-dimensional dual problem
\begin{equation}
\sup_{\mathbb{P}:\,\mathcal{D}_c(\mathbb{P},\mathbb{P}_n)\le\delta}
\mathbb{E}_{\mathbb{P}}\big[l(X^{(g)}, Y;\boldsymbol{\beta})\big]
=
\min_{\gamma \ge 0}
\left\{
\gamma\delta
+
\frac{1}{n}
\sum_{k=1}^n
\phi_\gamma(x_k^{(g)}, y_k; \boldsymbol{\beta})
\right\}.
\end{equation}

Substituting this expression into the original distributionally robust regression problem yields
\begin{equation}
\inf_{\boldsymbol{\beta} \in \mathbb{R}^{d}}
\sup_{\mathbb{P}:\,\mathcal{D}_c(\mathbb{P},\mathbb{P}_n)\le\delta}
\mathbb{E}_{\mathbb{P}}\big[l(X^{(g)}, Y; \boldsymbol{\beta})\big]
=
\inf_{\boldsymbol{\beta} \in \mathbb{R}^{d}}
\min_{\gamma\ge0}
\left\{
\gamma\delta
+
\frac{1}{n}\sum_{k=1}^n \phi_\gamma(x_k^{(g)}, y_k; \boldsymbol{\beta})
\right\}.
\end{equation}

We now apply the general notation $(X^{(g)},Y)$ and the associated loss $l(X^{(g)},Y;\boldsymbol{\beta})$ to the Gaussian graphical model framework. For a given target node $i \in \{1, \dots, p\}$, the generic response $Y$ corresponds to the variable $X_i$, and the generic predictor vector $X^{(g)}$ corresponds to the remaining variables $X_{-i}$. Accordingly, the loss function depends entirely on the joint random vector $X = (X_1, \dots, X_p)^\top$. With a slight abuse of notation, we write $l(X;\boldsymbol{\beta}^{(i)})$ when specializing to the node-wise neighborhood selection formulation. 

Previously, neighborhood selection was described from a macroscopic column-wise perspective using the data matrix $\mathbf{X} \in \mathbb{R}^{n \times p}$. In this section, to facilitate the distributionally robust analysis, we shift to an equivalent sample-wise representation. Let the $k$-th row of the data matrix $\mathbf{X}$ be denoted by $x_k^\top$, where $x_k \in \mathbb{R}^p$ represents the $k$-th full observation vector for $k \in \{1, \dots, n\}$. For our node-wise regression targeting node $i$, the response is the scalar $x_{k,i}$ (the $i$-th component of $x_k$), and the predictors are the $(p-1)$-dimensional vector $x_{k,-i}$ (the remaining components of $x_k$).

To express the node-wise loss more compactly, we introduce the augmented vectors for both the observations and the parameters:
\[
\bar{x}_{k,i} = (x_{k,-i}, x_{k,i}),
\qquad
\bar{\boldsymbol{\beta}}^{(i)}(K) = (-\boldsymbol{\beta}^{(i)}(K), 1),
\qquad
\bar{u}_i = (u_i, v_i).
\]
Here, $\boldsymbol{\beta}^{(i)}(K)$ denotes the population neighborhood selection coefficient associated with the $i$-th node in the Gaussian graphical model, which is uniquely determined by the precision matrix $K$ via $\beta^{(i)}_j(K) = -K_{ij}/K_{ii}$ for $j \neq i$, as reviewed in Section~\ref{sec:background}. We consider the node-wise quadratic loss
\[
l(x_{k,-i}, x_{k,i}; K)
=
\left( \big(\bar{\boldsymbol{\beta}}^{(i)}(K)\big)^{\top} \bar{x}_{k,i} \right)^2,
\]
and aggregate these node-wise objectives through scalarization to obtain the multi-task loss for the $k$-th sample:
\[
l(x_k;K)
=
\frac{1}{p}
\sum_{i=1}^{p}
\left( \big(\bar{\boldsymbol{\beta}}^{(i)}(K)\big)^{\top} \bar{x}_{k,i} \right)^2.
\]

Applying equation~\eqref{phi_gamma} to this scalarized loss yields the robustified node-wise function for the $k$-th observation:
\begin{equation}
\phi_{\boldsymbol{\gamma}}(x_k;K)
=
\sup_{\{\bar{u}_i\}_{i=1}^p}
\left\{
\frac{1}{p}
\sum_{i=1}^{p}
\left( \bar{\boldsymbol{\beta}}^{(i)}(K)^\top \bar{u}_i \right)^2
-
\frac{1}{p}
\sum_{i=1}^{p}
\gamma_i
\big\| \bar{x}_{k,i} - \bar{u}_i \big\|_{\kappa}^{2}
\right\},
\qquad
\bar{u}_i \in \mathbb{R}^{p}.
\end{equation}
Combining the above expression with the dual representation of Wasserstein DROP, we obtain the following closed-form equivalence for the distributionally robust multi-task learning problem:
\begin{equation} \label{DRO_multi}
\sup_{\mathbb{P}:\, \mathcal{D}_c(\mathbb{P}, \mathbb{P}_n) \le \delta}
\mathbb{E}_{\mathbb{P}}
\big[
l(X;K)
\big]
=
\frac{1}{p}
\sum_{i=1}^{p}
\left(
\sqrt{\mathrm{MSE}_n\big(\boldsymbol{\beta}^{(i)}(K)\big)}
+
\sqrt{\delta}\,
\big\| \bar{\boldsymbol{\beta}}^{(i)}(K) \big\|_{\kappa^{*}}
\right)^2,
\end{equation}
where $1/\kappa+1/\kappa^*=1$,  $\mathrm{MSE}_n(\boldsymbol{\beta}^{(i)}(K)) = \frac{1}{n} \sum_{k=1}^{n} \left( x_{k,i} - \boldsymbol{\beta}^{(i)}(K)^\top x_{k,-i} \right)^{2}$ denotes the empirical mean squared error associated with the $i$-th node-wise neighborhood selection regression, the parameter vector is augmented as $\bar{\boldsymbol{\beta}}^{(i)}(K) = (-\boldsymbol{\beta}^{(i)}(K), 1)$.

\begin{proposition}[Distributionally Robust Scalarized Multi-Task GGM] \label{proposition1}
Fix $\kappa \in (1,\infty)$ and let $\kappa^*$ satisfy $1/\kappa + 1/\kappa^* = 1$. Consider a Gaussian graphical model with precision matrix $K \succ 0$, and let $\{\boldsymbol{\beta}^{(i)}(K)\}_{i=1}^p$ denote the corresponding neighborhood selection coefficients. Assume that the distributional uncertainty set is defined via the second-order optimal transport discrepancy $\mathcal{D}_c$ induced by the $\ell_\kappa$-norm. Then
\[
\sup_{\mathbb{P}:\, \mathcal{D}_c(\mathbb{P}, \mathbb{P}_n) \le \delta}
\mathbb{E}_{\mathbb{P}}
\big[
l(X;K)
\big]
=
\frac{1}{p}
\sum_{i=1}^{p}
\left(
\sqrt{\mathrm{MSE}_n\big(\boldsymbol{\beta}^{(i)}(K)\big)}
+
\sqrt{\delta}\,
\big\| \bar{\boldsymbol{\beta}}^{(i)}(K) \big\|_{\kappa^{*}}
\right)^2,
\]
where $\mathrm{MSE}_n(\boldsymbol{\beta}^{(i)}(K)) = \frac{1}{n} \sum_{k=1}^{n} \left( x_{k,i} - \boldsymbol{\beta}^{(i)}(K)^\top x_{k,-i} \right)^{2}$ denotes the empirical mean squared error associated with the $i$-th node-wise neighborhood selection regression.
\end{proposition}

The detailed proof of this proposition is provided in Appendix. 
This formulation extends the foundational results of \citet{blanchet2019robust} from a single-task setting to a multi-task framework.


\subsection{The Proposed DROP Estimator}
Building on the distributionally robust formulation derived in the previous section, we now construct a tractable estimator for the precision matrix $K$. Minimizing the worst-case risk established in \eqref{DRO_multi} poses computational challenges due to the non-smooth and non-convex nature of the squared sum of square roots. To derive an efficient estimation procedure, we first make explicit the dependence of the regression coefficients on the precision matrix via the relationship $\beta^{(i)}_j(K) = -K_{ij}/K_{ii}$ for $j \neq i$.

Focusing on the $\ell_1$-norm ($\kappa^* = 1$) to induce sparsity on the off-diagonal elements, we substitute $\boldsymbol{\beta}^{(i)}(K)$ into the objective \eqref{DRO_multi}. Excluding the unpenalized intercept component from the augmented vector penalty, we can expand the squared term within the minimization problem as follows:
\begin{equation}
\label{eq:expansion}
\begin{split}
    & \sum_{i=1}^p \left( \sqrt{\mathrm{MSE}_{n}\big(\boldsymbol{\beta}^{(i)}(K)\big)} +\sqrt{\delta_i}\,\frac{\|K_{-i,i}\|_{1}}{K_{ii}} \right)^2 \\
    = & \sum_{i=1}^p \left( \mathrm{MSE}_{n}\big(\boldsymbol{\beta}^{(i)}(K)\big) + 2\sqrt{\delta_i}\, \frac{\sqrt{\mathrm{MSE}_{n}\big(\boldsymbol{\beta}^{(i)}(K)\big)}}{K_{ii}}\|K_{-i,i}\|_{1} + \frac{\delta_i}{K_{ii}^2} \|K_{-i,i}\|_{1}^2 \right),
\end{split}
\end{equation}
where $\mathrm{MSE}_n(\boldsymbol{\beta}^{(i)}(K)) = \frac{1}{n} \sum_{k=1}^{n} \left( x_{k,i} - \boldsymbol{\beta}^{(i)}(K)^\top x_{k,-i} \right)^{2}$ represents the empirical mean squared error for the $i$-th node-wise regression.

We propose to retain the first two terms in \eqref{eq:expansion} while omitting the third quadratic regularization term. This approximation is theoretically justified by the asymptotic behavior of the optimal transport radius. Typically, to ensure that the true data-generating distribution lies within the ambiguity set with high probability, the radius $\delta_i$ is chosen to decay with the sample size at a rate of $\mathcal{O}(n^{-1/2})$~\citep{blanchet2019robust}. Consequently, the third term in \eqref{eq:expansion}, which scales linearly with $\delta_i$, vanishes at a strictly faster rate asymptotically than the second cross-term, which scales with $\sqrt{\delta_i}$. The second term therefore constitutes the leading-order regularization effect induced by the distributional uncertainty.

By isolating this dominant term and introducing the tuning parameter $\lambda_i = 2\sqrt{\delta_i}$, we arrive at the proposed DROP estimator:
\begin{equation}
    \hat{K} = \arg\min_{K \succ 0} \sum_{i=1}^p \left\{ \mathrm{MSE}_{n}\big(\boldsymbol{\beta}^{(i)}(K)\big) + \lambda_i w^{(i)} \|K_{-i,i}\|_{1} \right\}, 
    \label{eq:drop_estimator}
\end{equation}
where the adaptive weights are defined as $w^{(i)} := \frac{\sqrt{\mathrm{MSE}_{n}(\boldsymbol{\beta}^{(i)}(K))}}{K_{ii}}$. This formulation naturally recovers an adaptively weighted $\ell_1$-penalized objective for each node. Crucially, rather than being specified arbitrarily, the weights $w^{(i)}$ are systematically derived from the distributionally robust framework, scaling adaptively with the node-wise prediction error and the diagonal precision matrix elements.

For computational tractability and notational convenience, we unify the regularization parameter as $\lambda$ and symmetrize the objective function. By expressing the empirical MSE in its vectorized $\ell_2$-norm form, the DROP estimator can be equivalently written as the solution to the following symmetric optimization problem:
\begin{equation} \label{eq:drop_symmetric}
  \hat{K} = \arg\min_{K \succ 0} \left\{ \sum_{i=1}^p \frac{1}{n} \big\| \mathbf{X}_i - \mathbf{X}_{-i} \boldsymbol{\beta}^{(i)}(K) \big\|_2^2 + \lambda \sum_{i<j} w_{ij}|K_{ij}| \right\},
\end{equation}
where $\mathbf{X}_i \in \mathbb{R}^n$ denotes the $i$-th column of the data matrix $\mathbf{X}$, $\mathbf{X}_{-i} \in \mathbb{R}^{n \times (p-1)}$ is the submatrix excluding the $i$-th column, and the symmetric adaptive weights are defined as
\begin{equation} \label{eq:drop_weights}
    w_{ij} := \frac{\sqrt{\mathrm{MSE}_{n}\big(\boldsymbol{\beta}^{(i)}(K)\big)}}{K_{ii}} + \frac{\sqrt{\mathrm{MSE}_{n}\big(\boldsymbol{\beta}^{(j)}(K)\big)}}{K_{jj}}.
\end{equation}

\section{Statistical Properties of DROP}
\label{sec:methodology}

Having introduced the DROP methodology, we now establish its theoretical foundations. In this section, we analyze the error bound properties of the foundational multi-task DROP framework that underpins our approach. To build the necessary theoretical tools, we first examine the standard single-task setting before extending our analysis to the joint multi-task problem.

\subsection{Single-Task Learning} \label{subsec:single_task}

Consider the standard linear regression model:
\begin{equation}\label{M1}
    \mathbf{y} = \mathbf{X}\boldsymbol{\beta}^* + \boldsymbol{\varepsilon},
\end{equation}
where $\mathbf{y} \in \mathbb{R}^n$ is the response vector, $\mathbf{X} \in \mathbb{R}^{n \times p}$ is the design matrix, $\boldsymbol{\beta}^* \in \mathbb{R}^p$ is the unknown sparse regression vector, and $\boldsymbol{\varepsilon} \in \mathbb{R}^n$ represents the random noise vector.

In high-dimensional scenarios, estimating a sparse coefficient vector $\boldsymbol{\beta}^*$ typically relies on $\ell_1$-penalized methods such as the Lasso~\citep{lounici2011oracle}. While the standard Lasso provides a well-established baseline (provided in Appendix), it offers no inherent robustness against data contamination. To immunize the estimation process against such distributional shifts, we cast the sparse regression problem within the Wasserstein DROP framework. Applying duality equivalence, the worst-case robust formulation integrates the robustness directly into the objective, yielding our single-task DROP estimator:
\begin{equation}\label{Est2}
    \hat{\boldsymbol{\beta}}_{\text{DROP}} = \arg\min_{\boldsymbol{\beta} \in \mathbb{R}^p}
    \left\{ \frac{1}{n}\|\mathbf{y}-\mathbf{X}\boldsymbol{\beta}\|_2^2
    + \frac{\lambda}{\sqrt{n}} \| \mathbf{y}-\mathbf{X}\boldsymbol{\beta}\|_2\|\boldsymbol{\beta}\|_1 \right\}.
\end{equation}

To establish the theoretical properties of both estimators, we require the following regularity conditions on the noise and the design matrix.

\begin{assumption}\label{assump1}
    Assume that $\boldsymbol{\varepsilon} \in \mathbb{R}^n$ is a random vector with i.i.d.\ $\mathcal{N}(\mathbf{0}, \sigma^2)$ Gaussian components, where $\sigma^2 > 0$. There exists a constant $c = c(\sigma^2, p) > 1$ and a constant $M_c$ such that
    \begin{equation*}
        \mathbb{P} \left( \frac{\| \boldsymbol{\varepsilon} \|_2}{\sqrt{n}} \leq c \right) \geq 1 - M_c.
    \end{equation*}
\end{assumption}

\begin{assumption}\label{assump2}
    Let $\mathbf{X} \in \mathbb{R}^{n \times p}$ be a fixed design matrix and denote its $j$-th column by $\mathbf{X}_{j}$, for $j=1, \ldots, p$. For a coefficient vector $\boldsymbol{\beta} \in \mathbb{R}^p$, define the active set $J(\boldsymbol{\beta})=\left\{j \in\{1, \ldots, p\}: \beta_j \neq 0\right\}$. Assume that $\boldsymbol{\varepsilon} \in \mathbb{R}^n$ is a random vector with i.i.d.\ $\mathcal{N}(\mathbf{0}, \sigma^2)$ Gaussian components, where $\sigma^2 > 0$. There exist a tuning parameter $\lambda > \frac{c}{\sqrt{n}}$ and a constant $c = c(\sigma^2, \mathbf{X}, p) > 0$ such that
    \begin{equation*}
        \mathbb{P} \left( \max_{1 \le j \le p} \frac{1}{n} \left| \mathbf{X}_j^\top \boldsymbol{\varepsilon} \right| \leq \frac{\lambda}{2} \right) \geq 1 - M_c.
    \end{equation*}
\end{assumption}

\begin{remark}
Assumption~\ref{assump1} states that the noise magnitude $\frac{1}{\sqrt{n}}\|\boldsymbol{\varepsilon}\|_2$ is bounded by a constant $c$ with probability at least $1-M_c$, ensuring that the overall error level is well controlled. Assumption~\ref{assump2} similarly requires that the empirical cross-correlation between the noise and each predictor, given by $\frac{1}{n}|\mathbf{X}_j^\top \boldsymbol{\varepsilon}|$, is strictly bounded by $\frac{\lambda}{2}$ with probability at least $1-M_c$, guaranteeing that random spurious correlations remain limited. Together, these assumptions ensure that the noise does not overwhelm the estimation procedure.
\end{remark}

As a baseline, the error bound guarantees of the standard Lasso estimator under these assumptions are well-established \citep{lounici2011oracle}. For completeness, we restate these foundational properties in the Appendix. Building upon this framework, we first establish the theoretical properties of the single-task DROP estimator. While our primary theoretical contribution lies in the multi-task framework presented in the subsequent section, analyzing the single-task setting serves as an essential theoretical stepping stone.

\begin{theorem}\label{thm1}
    Consider the model~\eqref{M1} and the estimator~\eqref{Est2}. Under Assumption~\ref{assump1} and Assumption~\ref{assump2}, with probability at least $1 - M_c$, the following statements hold: 
    \begin{enumerate}
        \item[(i)] \emph{(Basic Inequality).} \begin{equation}
            \frac{1}{n}\|\mathbf{y}-\mathbf{X} \hat{\boldsymbol{\beta}}_{\text{DROP}}\|_2^{2}+\frac{\lambda}{\sqrt{n}}\|\mathbf{y}-\mathbf{X} \hat{\boldsymbol{\beta}}_{\text{DROP}}\|_2 \|\hat{\boldsymbol{\beta}}_{\text{DROP}}\|_{1}
            \leq \frac{1}{n}\|\mathbf{y}-\mathbf{X} \boldsymbol{\beta}\|_2^{2}+\frac{\lambda}{\sqrt{n}}\|\mathbf{y}-\mathbf{X} \boldsymbol{\beta}\|_2 \|\boldsymbol{\beta}\|_{1} .
            \label{thm1state1}
        \end{equation}
        \item[(ii)] \emph{(Prediction Error Bound).} \begin{equation}
            \frac{1}{n}\|\mathbf{X} (\boldsymbol{\beta}^{*}-\hat{\boldsymbol{\beta}}_{\text{DROP}})\|_2^2 \leq 2c\lambda \sum_{j=1}^{p} |\beta^{*}_j-(\hat{\boldsymbol{\beta}}_{\text{DROP}})_j|+ 2c \lambda \sum_{j=1}^{p} |\beta^*_j| .
            \label{thm1state2}
        \end{equation}
        \item[(iii)] \emph{(Score Bound).} 
        \begin{equation}
        \begin{split}
            \frac{1}{n} \left| \big[\mathbf{X}^{\top} \mathbf{X}(\hat{\boldsymbol{\beta}}_{\text{DROP}}-\boldsymbol{\beta}^{*})\big]_j \right| 
            \leq & \; \frac{\lambda^{\frac{3}{2}}}{2} \sqrt{2c \sum_{j \in J(\boldsymbol{\beta}^*-\hat{\boldsymbol{\beta}}_{\text{DROP}})} |\beta^*_j - (\hat{\boldsymbol{\beta}}_{\text{DROP}})_j|} \\
            & + \frac{\lambda^{\frac{3}{2}}}{2} \sqrt{2c \sum_{j \in J(\boldsymbol{\beta}^*)} |\beta^*_j|} + \lambda c.
        \end{split}
        \label{thm1state3}
        \end{equation}
    \end{enumerate}
\end{theorem}

\begin{remark}\label{thm1_remark1}
    For Statement (iii) (Equation~\eqref{thm1state3}), if the minimum eigenvalue of the limit $\mathbf{X}^\top \mathbf{X}/n$ is positive and $\max\big(|\beta^*_j|, |(\hat{\boldsymbol{\beta}}_{\text{DROP}})_j|\big) \leq c^*$ with probability at least $1-M_{c^*}$, then $|(\hat{\boldsymbol{\beta}}_{\text{DROP}})_j-\beta^{*}_j| \leq \lambda^{3/2}c^{1/2}(2c^*|J(\boldsymbol{\beta}^*)|)^{1/2}+\frac{\lambda^2}{2}pc^*+\lambda c $ with probability at least $1-M_c-M_{c^*}$. When $\lambda=\mathcal{O}(n^{-1/2})$ and $p=o(n)$, we have $|(\hat{\boldsymbol{\beta}}_{\text{DROP}})_j-\beta^{*}_j|=o(1)$. When $\lambda=\mathcal{O}(n^{-1/2})$ and $p$ is fixed, $|(\hat{\boldsymbol{\beta}}_{\text{DROP}})_j-\beta^{*}_j|=\mathcal{O}(n^{-1/2})$.
\end{remark}

\begin{remark}\label{thm1_remark2}
    For Statement (ii) (Equation~\eqref{thm1state2}), if $c_1 I_p \leq \mathbf{X}^\top \mathbf{X}/n \leq c_2 I_p$ with probability at least $1-M_{c_1}$ and $\left\|\boldsymbol{\beta}^*\right\|_1 \neq 0$, then $\left\|\boldsymbol{\beta}^*-\hat{\boldsymbol{\beta}}_{\text{DROP}}\right\|_2 \leq o(\lambda \sqrt{p})+\mathcal{O}\left(\sqrt{\lambda\left\|\boldsymbol{\beta}^*\right\|_1}\right)$ with probability at least $1-M_c-M_{c_1}$. When $p \ll n$, we have $2c \lambda \sqrt{p}/{c_1} =  o(\lambda \sqrt{p})$ and $\sqrt{8c \lambda/c_1}=\mathcal{O}\left(\sqrt{\lambda\left\|\boldsymbol{\beta}^*\right\|_1}\right)$.
\end{remark}

The detailed proofs of Theorem~\ref{thm1} and its related remarks are provided in Appendix.

\subsection{Multi-Task Learning}
Building on the theoretical foundation of the single-task setting, we now extend our methodology to the multi-task learning scenario. For Gaussian Graphical Models, node-wise regression inherently poses a multi-task problem, where the reconstruction of the neighborhood for each node $i$ acts as a distinct but correlated task. 

Let $K^*$ denote the true precision matrix, and define $\boldsymbol{\beta}^{(i)}(K^*)$ as the target regression coefficient vector for the $i$-th task. Since the estimated regression coefficients collectively characterize the empirical precision matrix, we naturally denote the estimators parameterized by the precision matrix as $\boldsymbol{\beta}^{(i)}(K)$. The general multi-task learning model is formulated as:
\begin{equation}\label{eq:mt_model}
\begin{aligned}
\mathbf{y}_1 & = \mathbf{X}_1 \boldsymbol{\beta}^{(1)}(K^*) + \boldsymbol{\varepsilon}^{(1)}, \\
\mathbf{y}_2 & = \mathbf{X}_2 \boldsymbol{\beta}^{(2)}(K^*) + \boldsymbol{\varepsilon}^{(2)}, \\
& \vdots \\
\mathbf{y}_p & = \mathbf{X}_p \boldsymbol{\beta}^{(p)}(K^*) + \boldsymbol{\varepsilon}^{(p)}.
\end{aligned}
\end{equation}

The standard pooled Lasso estimator for this joint multi-task problem minimizes the empirical loss across all tasks:
\begin{equation}\label{muti_Est1}
\hat{\boldsymbol{\beta}}_{\text{Lasso}} = \arg\min_{\boldsymbol{\beta}} \frac{1}{p} \sum_{i=1}^{p} \left( \frac{1}{n}\|\mathbf{y}_i - \mathbf{X}_i \boldsymbol{\beta}^{(i)}(K)\|_2^2 + \lambda \|\boldsymbol{\beta}^{(i)}(K)\|_1 \right).
\end{equation}

However, to strengthen this multi-task estimation robust to distributional shifts and heavy-tailed dependencies, we reformulate the problem through the Wasserstein DROP framework. This yields the foundational multi-task DROP estimator, which serves as the theoretical building block for our DROP methodology developed in Section~\ref{sec:DROP}:
\begin{equation}\label{muti_Est2}
\hat{\boldsymbol{\beta}}_{\text{DROP}} = \arg\min_{\boldsymbol{\beta}} \frac{1}{p}\sum_{i=1}^p
\left(
\frac{1}{n}\|\mathbf{y}_i - \mathbf{X}_i \boldsymbol{\beta}^{(i)}(K)\|_2^2
+
\frac{\lambda}{\sqrt{n}}
\|\mathbf{y}_i - \mathbf{X}_i \boldsymbol{\beta}^{(i)}(K)\|_2
\|\boldsymbol{\beta}^{(i)}(K)\|_1
\right).  
\end{equation}

To theoretically analyze this estimator, we generalize our previous assumptions to accommodate task dependence and uniform sparsity across tasks.

\begin{assumption}\label{mutil_assump1}
(Multi-task noise; task dependence).
Assume that $\boldsymbol{E} = (\boldsymbol{\varepsilon}^{(1)},\dots,\boldsymbol{\varepsilon}^{(p)}) \in \mathbb{R}^{n\times p}$ is a random noise matrix. Each row of $\boldsymbol{E}$ follows a centered multivariate Gaussian distribution. In particular, the noise vectors $\boldsymbol{\varepsilon}^{(1)},\dots,\boldsymbol{\varepsilon}^{(p)}$ are allowed to be dependent across tasks. There exist constants $c>0$ and $C_{\varepsilon}\in(0,1)$, as well as a tuning parameter $\lambda > \frac{c}{\sqrt{n}}$, such that
\begin{equation*}
    \mathbb{P}\!\left(
    \frac{1}{p}\sum_{i=1}^p \frac{\|\boldsymbol{\varepsilon}^{(i)}\|_2^2}{n} \le c
    \ \textup{and}\
    \max_{1\le i\le p}\frac{\|\boldsymbol{\varepsilon}^{(i)}\|_2}{\sqrt{n}} \le C_{\varepsilon}
    \right) \ge 1 - M_c.
\end{equation*}
\end{assumption}

\begin{assumption}\label{mutil_assump2}
(Uniform score bound under task dependence).
    Let $\mathbf{X}_i \in \mathbb{R}^{n \times p}$ be a fixed design matrix for task $i=1,\dots,p$, and denote by $\mathbf{X}_{i,j}$ its $j$-th column. For a collection of coefficient vectors $\boldsymbol{\beta}(K)=(\boldsymbol{\beta}^{(1)}(K),\dots,\boldsymbol{\beta}^{(p)}(K))$, define the active task set $J(\boldsymbol{\beta}(K))=\{\, i\in\{1,\dots,p\}:\ \boldsymbol{\beta}^{(i)}(K) \neq \mathbf{0} \,\}$. There exists a tuning parameter $\lambda>0$ and a constant $c=c(\Sigma,\mathbf{X}_1,\dots,\mathbf{X}_p)>0$ such that
    \begin{equation*}
        \mathbb{P}\!\left(
        \max_{1\le i\le p}
        \left\|
        \frac{1}{n}\mathbf{X}_i^\top \boldsymbol{\varepsilon}^{(i)}
        \right\|_\infty
        \le \frac{\lambda}{2}
        \right)
        \ge 1 - M_c,
    \end{equation*}
    where $\|\cdot\|_\infty$ denotes the maximum absolute value over the coordinates $j$.
\end{assumption}

\begin{remark}
Assumption~\ref{mutil_assump1} follows the noise structure considered in~\citep{zhang2025multi}, where concentration bounds for Gaussian quadratic forms and sums of dependent chi-squared random variables are developed under task dependence.
\end{remark}

\begin{remark}

Assumptions~\ref{mutil_assump1} and~\ref{mutil_assump2} are multi-task extensions of the baseline conditions used for the standard Lasso \citep{lounici2011oracle}. In particular, Assumption~\ref{mutil_assump2} plays the same role as the classic score bound (restated in Appendix), while allowing for dependence across tasks.

We provide the rationale for why Assumption~\ref{mutil_assump2} holds under Assumption~\ref{mutil_assump1} when the tasks are dependent. For each task $i$ and coordinate $j$, the score term $\frac{1}{n}(\mathbf{X}_i^\top \boldsymbol{\varepsilon}^{(i)})_j = \frac{1}{n}\sum_{k=1}^n (\mathbf{X}_i)_{kj}(\boldsymbol{\varepsilon}^{(i)})_k$ is a centered Gaussian random variable conditional on $\mathbf{X}_i$, with variance of order $\mathcal{O}(1/n)$ under standard normalization of the columns of $\mathbf{X}_i$. Unlike the independent-task setting, the random variables $\{\mathbf{X}_i^\top \boldsymbol{\varepsilon}^{(i)}\}_{i=1}^p$ are not independent across $i$ due to the cross-task covariance structure induced by $\Sigma$.

Following the approach of~\citep{zhang2025multi}, the task dependence can be represented by a dependency graph whose vertices correspond to tasks and whose edges encode the nonzero correlations implied by $\Sigma$. A standard graph-coloring argument partitions the set of tasks into a finite number of groups such that, within each group, the corresponding score variables are independent. Applying Gaussian tail bounds within each group and combining the results across groups yields that, for a suitable choice of $\lambda$, $\mathbb{P} \big( \max_{1\le i\le p} \| \frac{1}{n}\mathbf{X}_i^\top \boldsymbol{\varepsilon}^{(i)} \|_\infty > \frac{\lambda}{2} \big) \le M_c$, which establishes Assumption~\ref{mutil_assump2}.
\end{remark}

\begin{assumption}\label{assump:RE}
(Restricted eigenvalue condition).
Let $\mathbf{X}_i \in \mathbb{R}^{n \times p}$ be fixed design matrices for tasks $i=1,\dots,p$. Let $K^*$ be the target parameter and define $S_i := \mathrm{supp}\big(\boldsymbol{\beta}^{(i)}(K^*)\big)$ and $s_i := |S_i|$. There exists a constant $\psi > 0$ such that for all collections $\mathbf{h} = (\mathbf{h}^{(1)}, \dots, \mathbf{h}^{(p)})$ satisfying the cone constraint
\begin{equation}\label{eq:cone_assump}
\sum_{i=1}^p \|\mathbf{h}^{(i)}_{S_i^c}\|_1 \le 3 \sum_{i=1}^p \|\mathbf{h}^{(i)}_{S_i}\|_1,
\end{equation}
the following inequality holds:
\begin{equation}\label{eq:RE_assump}
\frac{1}{p}\sum_{i=1}^p \frac{1}{n} \|\mathbf{X}_i \mathbf{h}^{(i)}\|_2^2 \ge \psi \cdot \frac{1}{p} \sum_{i=1}^p \|\mathbf{h}^{(i)}\|_2^2.
\end{equation}
\end{assumption}

\begin{remark}\label{rem:s_vs_J}
(Relationship between $S_i, s_i$ and $J(\boldsymbol{\beta})$).
For each task $i$, $S_i=\mathrm{supp}\big(\boldsymbol{\beta}^{(i)}(K^*)\big)$ is the within-task support and $s_i=|S_i|$ measures the within-task sparsity. The active task set $J\big(\boldsymbol{\beta}(K^*)\big)=\{\,i \in \{1,\dots,p\}:\boldsymbol{\beta}^{(i)}(K^*)\neq \mathbf{0}\,\}$ captures task-level sparsity, i.e., the number of active tasks. Clearly, if $i \notin J\big(\boldsymbol{\beta}(K^*)\big)$, then $\boldsymbol{\beta}^{(i)}(K^*) = \mathbf{0}$, which implies $S_i = \varnothing$ and $s_i = 0$. Consequently, the total number of nonzero coefficients across all tasks is given by $\sum_{i=1}^p s_i = \sum_{i \in J(\boldsymbol{\beta}(K^*))} s_i$.
\end{remark}

\begin{assumption}\label{assump:uniform_sparsity}
(Uniform within-task sparsity).
There exists an integer $s_{\max} \ge 1$ such that
\begin{equation*}
    s_i = \big|\mathrm{supp}(\boldsymbol{\beta}^{(i)}(K^*))\big| \le s_{\max}, \qquad \forall i=1,\dots,p.
\end{equation*}
\end{assumption}

\begin{remark}\label{rem:s_max_role}
Assumption~\ref{assump:uniform_sparsity} enforces a uniform upper bound on the within-task sparsity levels. It implies that the average sparsity $\frac{1}{p}\sum_{i=1}^p s_i$ appearing in Lemma~\ref{lem2}(iii) is bounded by $s_{\max}$, which yields a task-uniform $\ell_2$ estimation error bound independent of the number of tasks $p$.
\end{remark}

We first establish the baseline bounds for the standard multi-task Lasso estimator. The following lemma adapts the classic oracle inequalities established by \citet{lounici2011oracle} for the multi-task framework to explicitly accommodate our task-dependent assumptions.

\begin{lemma}\label{lem2}
Consider the multi-task model \eqref{eq:mt_model} and the standard Lasso estimator $\hat{\boldsymbol{\beta}}_{\text{Lasso}}$ defined in \eqref{muti_Est1}. Let $\boldsymbol{\beta}(K^*) = (\boldsymbol{\beta}^{(1)}(K^*),\dots,\boldsymbol{\beta}^{(p)}(K^*))$ be the true parameters and denote the estimator components by $\hat{\boldsymbol{\beta}}_{\text{Lasso}} = (\boldsymbol{\beta}^{(1)}(\hat K),\dots,\boldsymbol{\beta}^{(p)}(\hat K))$. Define the active task set $J\big(\boldsymbol{\beta}(K^*)\big) = \{ i\in\{1,\dots,p\}:\ \boldsymbol{\beta}^{(i)}(K^*)\neq \mathbf{0} \}$. Under Assumptions~\ref{mutil_assump1} and~\ref{mutil_assump2}, with probability at least $1-M_c$, the following statements hold:
\begin{enumerate}
    \item[(i)] \emph{(Basic Inequality).} 
    \begin{equation}
    \begin{split}
        & \frac{1}{p}\sum_{i=1}^p \left( \frac{1}{n}\|\mathbf{y}_i - \mathbf{X}_i \boldsymbol{\beta}^{(i)}(\hat K)\|_2^{2} + \lambda\|\boldsymbol{\beta}^{(i)}(\hat K)\|_{1} \right) \\
        & \le \frac{1}{p}\sum_{i=1}^p \left( \frac{1}{n}\|\mathbf{y}_i - \mathbf{X}_i \boldsymbol{\beta}^{(i)}(K^*)\|_2^{2} + \lambda\|\boldsymbol{\beta}^{(i)}(K^*)\|_{1} \right).
    \end{split}
    \label{lemma2state1}
    \end{equation}
    \item[(ii)] \emph{(Prediction Error Bound).} \begin{equation}
        \frac{1}{p}\sum_{i=1}^p \frac{1}{n}\|\mathbf{X}_i(\boldsymbol{\beta}^{(i)}(K^*)-\boldsymbol{\beta}^{(i)}(\hat K))\|_2^{2} \le \frac{2\lambda}{p} \sum_{i\in J(\boldsymbol{\beta}(K^*))} \min\!\Big( \|\boldsymbol{\beta}^{(i)}(K^*)\|_1,\ \|\boldsymbol{\Delta}^{(i)}\|_1 \Big).
        \label{lemma2state2}
        \end{equation}
    \item[(iii)] \emph{($\ell_2$ Error Bound).} Assume further that Assumption~\ref{assump:RE} holds with constant $\psi > 0$. Let $S_i=\mathrm{supp}\big(\boldsymbol{\beta}^{(i)}(K^*)\big)$ and $s_i=|S_i|$, and define $\boldsymbol{\Delta}^{(i)}:=\boldsymbol{\beta}^{(i)}(\hat K)-\boldsymbol{\beta}^{(i)}(K^*)$. Then, on the event in Assumption~\ref{mutil_assump2},
        \begin{equation}\label{lemma2state3}
        \frac{1}{p}\sum_{i=1}^p \|\boldsymbol{\beta}^{(i)}(\hat K)-\boldsymbol{\beta}^{(i)}(K^*)\|_2^2 \le \frac{64\,\lambda^2}{\psi^2}\, s_{\max}.
        \end{equation}
\end{enumerate}
\end{lemma}

The detailed proof of Lemma~\ref{lem2} is provided in Appendix. Finally, we present the theoretical guarantees for our foundational multi-task DROP framework.

\begin{theorem}\label{thm:mt_dro}
Consider the multi-task model \eqref{eq:mt_model} and the multi-task DROP estimator $\hat{\boldsymbol{\beta}}_{\text{DROP}}$ defined in \eqref{muti_Est2}. Let $\boldsymbol{\beta}(K^*) = (\boldsymbol{\beta}^{(1)}(K^*),\dots,\boldsymbol{\beta}^{(p)}(K^*))$ and let the resulting minimizer be $\hat{\boldsymbol{\beta}}_{\text{DROP}} = (\boldsymbol{\beta}^{(1)}(\hat K),\dots,\boldsymbol{\beta}^{(p)}(\hat K))$. Let $\boldsymbol{\Delta}^{(i)} := \boldsymbol{\beta}^{(i)}(\hat K) - \boldsymbol{\beta}^{(i)}(K^*)$ for $i=1,\dots,p$, and define the active task set $J\big(\boldsymbol{\beta}(K^*)\big) = \{ i\in\{1,\dots,p\}:\ \boldsymbol{\beta}^{(i)}(K^*) \neq \mathbf{0} \}$. Under Assumptions~\ref{mutil_assump1} and~\ref{mutil_assump2}, with probability at least $1-M_c$, the following statements hold:

\begin{enumerate}
    \item[(i)] \emph{(Basic Inequality).} For any $\boldsymbol{\beta} = (\boldsymbol{\beta}^{(1)}(K),\dots,\boldsymbol{\beta}^{(p)}(K))$,
        \begin{equation}
        \begin{aligned}
        & \sum_{i=1}^p \left( \frac{1}{n}\|\mathbf{y}_i - \mathbf{X}_i \boldsymbol{\beta}^{(i)}(\hat K)\|_2^2 + \frac{\lambda}{\sqrt{n}} \|\mathbf{y}_i - \mathbf{X}_i \boldsymbol{\beta}^{(i)}(\hat K)\|_2 \|\boldsymbol{\beta}^{(i)}(\hat K)\|_1 \right) \\ 
        & \le \sum_{i=1}^p \left( \frac{1}{n}\|\mathbf{y}_i - \mathbf{X}_i \boldsymbol{\beta}^{(i)}(K^*)\|_2^2 + \frac{\lambda}{\sqrt{n}} \|\mathbf{y}_i - \mathbf{X}_i \boldsymbol{\beta}^{(i)}(K^*)\|_2 \|\boldsymbol{\beta}^{(i)}(K^*)\|_1 \right).  
        \end{aligned}
        \label{thm:mt_dro_i}
        \end{equation}

    \item[(ii)] \emph{(Prediction Error Bound).} The prediction error satisfies
        \begin{equation}\label{eq:thm2_ii_final}
        \sum_{i=1}^p \frac{1}{n}\|\mathbf{X}_i \boldsymbol{\Delta}^{(i)}\|_2^2 \le \lambda(1+C_{\varepsilon}) \sum_{i=1}^p \|\boldsymbol{\Delta}^{(i)}\|_1 + \lambda \sum_{i=1}^p \frac{\|\mathbf{X}_i \boldsymbol{\Delta}^{(i)}\|_2}{\sqrt{n}} \|\boldsymbol{\beta}^{(i)}(\hat K)\|_1.
        \end{equation}

    \item[(iii)] \emph{(Natural Cone Constraint).} Assume further that Assumption~\ref{assump:RE} holds. Then, with probability at least $1-M_c$, the error $\boldsymbol{\Delta} = (\boldsymbol{\Delta}^{(1)},\dots,\boldsymbol{\Delta}^{(p)})$ satisfies the cone constraint
        \begin{equation}\label{eq:thm2_iii_cone_final}
        \sum_{i=1}^p \|\boldsymbol{\Delta}^{(i)}\|_1 \le \frac{2}{1-C_{\varepsilon}} \sum_{i=1}^p \|\boldsymbol{\Delta}^{(i)}_{S_i}\|_1 \;+\; \frac{1}{1-C_{\varepsilon}} \sum_{i=1}^p \frac{\|X_i\boldsymbol{\Delta}^{(i)}\|_2}{\sqrt{n}} \|\boldsymbol{\beta}^{(i)}(\hat K)\|_1 .
        \end{equation}
\end{enumerate}
\end{theorem}

The theoretical results in Theorem~\ref{thm:mt_dro} characterize the error bound properties of the robust multi-task framework. Statement (i) provides a deterministic basic inequality bounding the empirical loss. Statement (ii) establishes the prediction error bound, showing that the error explicitly scales with the regularization parameter $\lambda$ and the maximum noise perturbation $C_{\varepsilon}$. Finally, Statement (iii) proves that the estimation error belongs to a restricted cone, even under cross-task dependence and high dimensionality. This cone constraint is essential, as it guarantees that the error vector satisfies the restricted eigenvalue condition (Assumption~\ref{assump:RE}), which allows us to translate the prediction error bound into parameter estimation consistency.

\begin{remark}\label{rem:mt_dro_refined_drop_pred}
(Refined bounds under bounded solution size; dropping the prediction term).
Consider Statements~\textup{(ii)} and~\textup{(iii)} of Theorem~\ref{thm:mt_dro}. Suppose that there exists a constant $c_{\beta}>0$ such that $\lambda\max_{1\le i\le p}\|\boldsymbol{\beta}^{(i)}(\hat K)\|_1 \le c_{\beta}$ with probability at least $1-M_{c_{\beta}}$. Then, on the event in Assumption~\ref{mutil_assump1}, we have $\sum_{i=1}^p \frac{1}{n}\|\mathbf{X}_i\boldsymbol{\Delta}^{(i)}\|_2^2 \le 2\lambda(1+C_{\varepsilon}) \sum_{i=1}^p \|\boldsymbol{\Delta}^{(i)}\|_1 + 2c_{\beta}^2$, and the cone constraint in Theorem~\ref{thm:mt_dro} \textup{(iii)} further yields
\begin{equation*}
\sum_{i=1}^p \|\boldsymbol{\Delta}^{(i)}\|_1 \le \frac{4}{1-C_{\varepsilon}} \sum_{i=1}^p \|\boldsymbol{\Delta}^{(i)}_{S_i}\|_1 \;+\; \left( \frac{2(1+C_{\varepsilon})}{(1-C_{\varepsilon})^{2}} + \frac{2\sqrt{2}}{1-C_{\varepsilon}} \right)\frac{c_{\beta}^2}{\lambda},
\end{equation*}
with probability at least $1-M_c - M_{c_{\beta}}$.
\end{remark}

The detailed proof of Theorem~\ref{thm:mt_dro} is provided in Appendix. These error bound guarantees formally establish the statistical optimality of the robust multi-task framework that underpins the DROP estimator introduced in Section~\ref{sec:DROP}, demonstrating its rigorous control over both prediction and parameter estimation errors even under complex cross-task dependence and high dimensionality.

\section{Computational Algorithm}
\label{sec:algorithm}

The complete estimation procedure for the proposed DROP method is summarized in Algorithm~\ref{alg: DROP}. The algorithm proceeds in four main stages: data transformation, robust initialization, alternating optimization via coordinate descent, and final support recovery. In the following subsections, we provide detailed descriptions and theoretical justifications for each component.

\begin{algorithm}[!htbp]
\small
\caption{Distributionally Robust Optimization for Precision matrices (DROP)}
\label{alg: DROP}
\begin{algorithmic}[1]
\REQUIRE Data matrix $\mathbf{X} \in \mathbb{R}^{n \times p}$, regularization parameter $\lambda$, convergence tolerance $\epsilon = 10^{-4}$
\ENSURE Precision matrix $\hat{K}$ and adjacency matrix $\hat{A}$

\STATE \textbf{Step 1: Data Transformation (NPN)}
\STATE Apply rank-based inverse normal transformation to each variable $j$: 
\STATE \quad $\tilde{\mathbf{X}}_{kj} \gets \Phi^{-1}\left(\frac{\text{rank}(\mathbf{X}_{kj}) - 0.5}{n}\right)$ for $k=1,\dots,n$ and $j=1,\dots,p$

\STATE \textbf{Step 2: Robust Initialization}
\STATE Compute covariance estimator $\hat{\Sigma}_{\text{init}}$ using \texttt{CovTools}
\STATE Initialize $\hat{K}^{(0)} \gets (\hat{\Sigma}_{\text{init}})^{-1}$
\STATE Initialize $\text{iter} \gets 1$

\STATE \textbf{Step 3: Alternating Optimization}
\REPEAT
    \STATE Update symmetric adaptive weights $w_{ij}$ based on current $\hat{K}^{(\text{iter}-1)}$ via Eq.~\eqref{eq:drop_weights}
    \FOR{each pair $(i,j)$ with $1 \le i < j \le p$}
        \STATE Update $\hat{K}_{ij}$ using the closed-form coordinate descent solution in Eq.~\eqref{eq:final_update}
        \STATE $\hat{K}_{ji} \gets \hat{K}_{ij}$
    \ENDFOR
    \STATE $\text{iter} \gets \text{iter} + 1$
\UNTIL{convergence criterion $\|\hat{K}^{(\text{iter})} - \hat{K}^{(\text{iter}-1)}\|_1 < \epsilon$ is met}

\STATE \textbf{Step 4: Support Recovery}
\STATE $\hat{A}_{ij} \gets \mathbb{I}(|\hat{K}_{ij}| > 0)$
\RETURN $\hat{K}$, $\hat{A}$
\end{algorithmic}
\end{algorithm}

\subsection{Data Transformation via Nonparanormal}
The first stage of our algorithm involves transforming the input data to enhance robustness against heavy-tailed distributions and outliers. Following the nonparanormal (NPN) framework~\citep{liu2009nonparanormal}, we apply a rank-based inverse normal transformation to each variable. Specifically, for the $j$-th variable, the transformed observation for the $k$-th sample is computed as:
\begin{equation}
    \tilde{\mathbf{X}}_{kj} = \Phi^{-1}\left(\frac{\text{rank}(\mathbf{X}_{kj}) - 0.5}{n}\right), \quad \text{for } k = 1, \ldots, n,
    \label{eq:npn_transform}
\end{equation}
where $\Phi^{-1}$ denotes the inverse standard normal cumulative distribution function, and $\text{rank}(\mathbf{X}_{kj})$ is the rank of $\mathbf{X}_{kj}$ among the $n$ observations of the $j$-th variable.

This transformation offers several methodological advantages. First, it is distribution-free, accommodating any continuous marginal distribution without requiring parametric assumptions. Second, by relying solely on rank information, it effectively truncates the influence of extreme values, providing intrinsic protection against outliers. Theoretically, under the semiparametric Gaussian copula model~\citep{liu2012high}, the transformed data matrix $\tilde{\mathbf{X}}$  approximately follows a multivariate Gaussian distribution. This justifies the subsequent application of our DROP estimator while preserving the underlying conditional independence structure. In all subsequent optimization steps, we use the centered transformed data matrix $\tilde{\mathbf{X}}$ as the input.


\subsection{Robust Initialization}

To address the computational challenges in high-dimensional settings ($p > n$), we require a well-conditioned starting point for the iterative optimization procedure. Since the standard empirical covariance matrix is singular when $p > n$, directly inverting it to obtain an initial precision matrix is infeasible. 

As indicated in \textbf{Step 2} of Algorithm~\ref{alg: DROP}, we employ a regularized robust covariance estimator, denoted by $\hat{\Sigma}_{\text{init}}$~\citep{covtools2025}. The inverse of this well-conditioned estimator, $K^{(0)} = (\hat{\Sigma}_{\text{init}})^{-1}$, serves as our warm start. This strategy avoids singular matrix inversions, ensures numerical stability, and accelerates the convergence of the subsequent coordinate descent loop.

\subsection{Coordinate-wise Update Rule}

The core of our optimization procedure (\textbf{Step 3}) relies on the coordinate descent method. For a given pair $(i, j)$ with $i < j$, we update the precision matrix element $K_{ij}$ by minimizing the objective function while holding all other parameters fixed.

Let $\tilde{\mathbf{X}} \in \mathbb{R}^{n \times p}$ denote the transformed data matrix. We define the gradient component $S_0$ (aggregating terms independent of $K_{ij}$) and the curvature component $S_1$ (coefficient of $K_{ij}$) derived from the data-fitting term as:
\begin{align}
    S_{0} &= 2 \tilde{\mathbf{X}}_{i}^{\top} \tilde{\mathbf{X}}_{j} \left(\frac{1}{K_{i i}}+\frac{1}{K_{j j}}\right) + \frac{2}{K_{i i}^{2}} \tilde{\mathbf{X}}_{j}^{\top} \sum_{m \neq i,j} K_{i m} \tilde{\mathbf{X}}_{m} + \frac{2}{K_{j j}^{2}} \tilde{\mathbf{X}}_{i}^{\top} \sum_{m \neq i,j} K_{j m} \tilde{\mathbf{X}}_{m}, \label{eq:S0_def} \\
    S_{1} &= \frac{2}{K_{i i}^{2}} \tilde{\mathbf{X}}_{j}^{\top} \tilde{\mathbf{X}}_{j}+\frac{2}{K_{j j}^{2}} \tilde{\mathbf{X}}_{i}^{\top} \tilde{\mathbf{X}}_{i}. \label{eq:S1_def}
\end{align}

By analyzing the subgradient optimality condition, we obtain the closed-form update rule via the soft-thresholding operator. The optimal $\hat{K}_{ij}$ is given by:
\begin{equation}
\hat{K}_{i j} = 
\begin{cases}
\frac{-S_{0}+n \lambda w_{ij}}{S_{1}}, & \text{if } S_{0} > n \lambda w_{ij}, \\
\frac{-S_{0}-n \lambda w_{ij}}{S_{1}}, & \text{if } S_{0} < -n \lambda w_{ij}, \\
0, & \text{if } \left|S_{0}\right| \le n \lambda w_{ij}.
\end{cases}
\label{eq:final_update}
\end{equation}

This explicit solution guarantees the descent property of the objective function at each iteration. Detailed derivations of the partial derivatives and the stationarity conditions are relegated to Appendix.

\subsection{Model Selection}

The selection of the regularization parameter $\lambda$ dictates the sparsity of the estimated graph and is critical for high-dimensional performance. While Cross-Validation (CV) is a standard approach, it can be computationally prohibitive and prone to overfitting in high-dimensional settings~\citep{liu2010stability}. Information criteria such as AIC and BIC offer efficient alternatives; however, they tend to select overly dense graphs with excessive false positive edges when $p$ grows with $n$~\citep{foygel2010extended}. Alternatively, the Stability Approach to Regularization Selection (StARS)~\citep{liu2010stability} effectively controls the false discovery rate but remains computationally intensive due to repeated subsampling.

To address the consistency limitations of standard information criteria and the computational burden of resampling methods, we adopt the Extended Bayesian Information Criterion (EBIC)~\citep{chen2008extended, foygel2010extended}. For our DROP framework, we adapt the EBIC by replacing the standard Gaussian log-likelihood with the sum of node-wise mean squared errors, aligning completely with our scalarized multi-task objective. Specifically, we select the optimal $\lambda$ from a predefined grid by minimizing:
\begin{equation}
    \text{EBIC}_{\gamma_{\text{ebic}}}(\lambda) = n \log\left( \sum_{i=1}^p \mathrm{MSE}_{n}\big(\boldsymbol{\beta}^{(i)}(\hat{K}_\lambda)\big) \right) + |\hat{\mathcal{E}}_\lambda| \log n + 4\gamma_{\text{ebic}} |\hat{\mathcal{E}}_\lambda| \log p,
    \label{eq:ebic}
\end{equation}
where $\mathrm{MSE}_{n}\big(\boldsymbol{\beta}^{(i)}(\hat{K}_\lambda)\big)$ represents the empirical mean squared error for the $i$-th node-wise regression evaluated at the estimated precision matrix $\hat{K}_\lambda$, and $|\hat{\mathcal{E}}_\lambda|$ denotes the number of non-zero edges in the estimated graph. The criterion is parameterized by $\gamma_{\text{ebic}} \in [0,1]$. Following standard practice for high-dimensional sparse graph recovery, the parameter $\gamma_{\text{ebic}}$ is typically set to $0.5$ to ensure selection consistency~\citep{foygel2010extended}.


\section{Simulation Studies}
\label{sec:simulations}

We conduct comprehensive simulation studies to evaluate the performance of DROP compared to 10 competing methods across diverse graph structures and contamination scenarios. Our experimental design systematically varies dimensionality, sample size, graph type, and data contamination to assess method performance under realistic conditions.

\subsection{Simulation Settings and Performance Evaluation}
\label{subsec:design}

\begin{figure}[H]
    \centering
    \includegraphics[width=\textwidth]{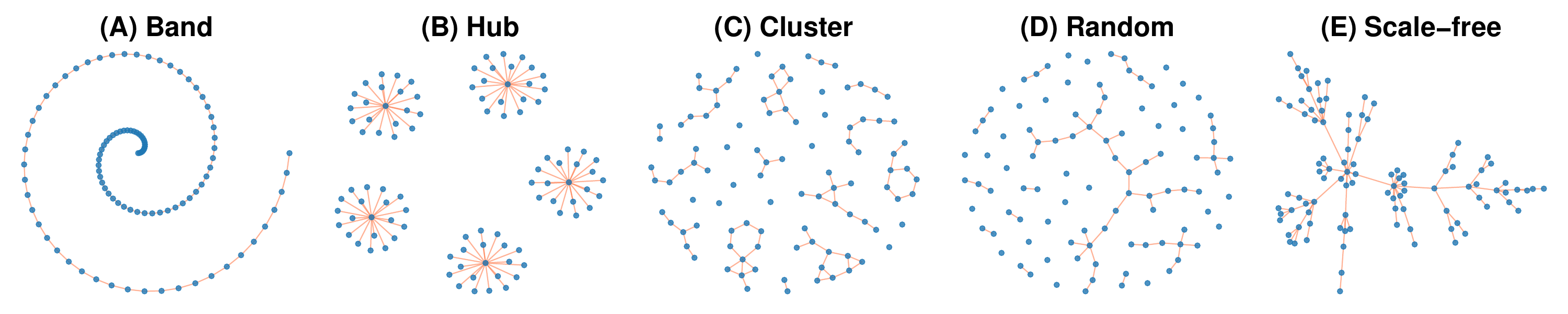}
    \caption{\footnotesize Illustration of the five graph structures with $p = 100$ nodes. (A) Band graph; (B) Hub graph; (C) Cluster graph; (D) Random graph; (E) Scale-free graph.}
    \label{fig:graph_structures}
\end{figure}

We generate data from a multivariate Gaussian distribution $\mathcal{N}(\mathbf{0}, \Sigma)$ across five distinct network structures. To visualize the complexity of these structures, we provide illustrations with $p=100$ nodes in Figure \ref{fig:graph_structures}, generated using the \texttt{huge} package \citep{zhao2012huge}.

\begin{enumerate}
    \item \textbf{Band graph:} A chain-like structure where vertices $i$ and $j$ are connected if $|i-j| \leq 2$, simulating the decay of correlation in time series data (Figure \ref{fig:graph_structures}(A)).
    \item \textbf{Hub graph:} The vertices are organized into disjoint stars. The center node of each star is connected to all other nodes in the cluster, creating high degree heterogeneity (Figure \ref{fig:graph_structures}(B)).
    \item \textbf{Cluster graph:} Vertices are divided into communities with a higher probability of within-cluster edges compared to between-cluster edges, resulting in a modular graph type (Figure \ref{fig:graph_structures}(C)).
    \item \textbf{Random graph:} Edges are assigned independently with probability $0.1$, following the Erdős-Rényi model \citep{erdos1959random} (Figure \ref{fig:graph_structures}(D)).
    \item \textbf{Scale-free graph:} Generated using the Barabási-Albert model \citep{barabasi1999emergence}, characterized by a power-law degree distribution (Figure \ref{fig:graph_structures}(E)).
\end{enumerate}

To assess the robustness of different methods, we consider three contamination schemes that represent common departures from Gaussianity. For all contamination scenarios, the contamination rate is set to $r = 0.1$.

\begin{enumerate}
    \item \textbf{Clean (no contamination):} Data are generated from the true Gaussian graphical model $\mathbf{x}_k \sim \mathcal{N}(\mathbf{0}, \mathbf{\Sigma})$ for $k = 1, \ldots, n$. This scenario serves as a baseline to verify that robust methods do not sacrifice efficiency when the Gaussian assumption holds.

    \item \textbf{Cauchy (Heavy-tailed contamination):} To evaluate robustness against extreme heavy-tailed noise, we consider an $r$-contamination setting in which a subset of observations is corrupted by additive Cauchy noise. Specifically, $10\%$ of randomly selected samples are replaced by
    \[
    \mathbf{x}_k \leftarrow \mathbf{x}_k + (\boldsymbol{\varepsilon}_{\text{cauchy}})_k,
    \quad (\boldsymbol{\varepsilon}_{\text{cauchy}})_k \sim \text{Cauchy}(\mathbf{0}, 5\mathbf{I}_p).
    \]
    This setup corresponds to a Huber-type contamination model with a severely heavy-tailed contamination distribution \citep {LohTan2018,AvellaMedina2018}.

    \item \textbf{Leverage (Leverage point contamination):} To model leverage-type outliers that preserve the underlying dependence structure but exhibit substantially inflated variability, we replace $10\%$ of the observations with samples drawn from a Gaussian distribution with an inflated covariance matrix,
    \[
    \mathbf{x}_k \sim \mathcal{N}(\mathbf{0}, 100\mathbf{\Sigma}).
    \]
    This contamination mechanism generates high-leverage points that preserve the correlation structure of the uncontaminated data, while having substantially larger variance. Such points can have a strong impact on covariance and precision matrix estimation~\citep{Hirose2017,onizuka2025robust}.
\end{enumerate}

We systematically evaluate DROP against a comprehensive suite of 10 state-of-the-art estimation methods, detailed in Table \ref{tab:competing_methods}. These baselines include standard non-robust estimators (\texttt{Glasso}, \texttt{MB}), non-convex penalizations (\texttt{SCAD}, \texttt{MCP}), alternative optimization strategies (\texttt{CLIME}, \texttt{TIGER}, \texttt{Scaled Lasso}), and traditional rank-based robust transformations (\texttt{NPN}, \texttt{NPN-Kendall}, \texttt{NPN-Spearman}).

\begin{table}[H]
\centering
\caption{Competing methods included in the simulation studies}
\label{tab:competing_methods}
\footnotesize
\begin{tabular}{@{}llp{6.5cm}@{}}
\toprule
Method & Package & Key Features \\
\midrule
Glasso & \texttt{huge} & Standard Gaussian graphical model with $\ell_1$ penalty \citep{friedman2008sparse} \\
MB & \texttt{huge} & Neighborhood selection via Lasso \citep{meinshausen2006high} \\
Scaled Lasso & \texttt{scalreg} & Joint estimation of precision matrix and noise level \citep{sun2012scaled} \\
SCAD & \texttt{GGMncv} & Nonconvex SCAD penalty to reduce estimation bias \citep{fan2001variable} \\
MCP & \texttt{GGMncv} & Minimax concave penalty \citep{zhang2010nearly} \\
TIGER & \texttt{flare} & Tuning-free estimator via column-wise inverse regression \citep{liu2017tiger} \\
CLIME & \texttt{flare} & Constrained $\ell_1$ minimization for inverse matrix estimation \citep{cai2011constrained} \\
\midrule
\multicolumn{3}{@{}l}{\textit{Rank-based Nonparanormal Methods (Robust Baselines)}} \\
NPN & \texttt{huge} & Rank-based normal score transformation + Glasso \citep{liu2009nonparanormal, zhao2012huge} \\
NPN-Kendall & Custom + \texttt{huge} & Kendall's $\tau$ correlation matrix + Glasso (The ``Skeptic'') \citep{liu2009nonparanormal} \\
NPN-Spearman & Custom + \texttt{huge} & Spearman's $\rho$ correlation matrix + Glasso \citep{xue2012regularized} \\
\bottomrule
\end{tabular}
\end{table}

To thoroughly investigate both finite-sample properties and algorithmic scalability, the experiments are divided into two main simulation regimes based on settings established in recent literature \citep{li2021penalized, lingjaerde2024scalable}:

\begin{enumerate}
    \item \textbf{Moderate-dimensional regime ($p \le 30$):} This setting assesses the robust structural recovery of the estimators under finite samples. We consider combinations of dimension $p \in \{10, 20, 30\}$ and sample size $n \in \{250, 500, 1000\}$. In this regime, DROP is rigorously compared against all 10 competing methods, yielding a comprehensive evaluation across $3 \times 3 \times 5 \times 3 = 135$ distinct experimental configurations.
    
    \item \textbf{High-dimensional regime ($p \in \{100, 250\}$):} To demonstrate the practical scalability of our method, we extend the evaluation to larger network structures. Specifically, we set $n \in \{100, 1000\}$ for $p=100$, and $n \in \{500, 5000\}$ for $p=250$. The latter setting is specifically designed to parallel the scale of our subsequent real data application. Due to the heavy computational burden in high dimensions, we filter the competing methods and compare DROP exclusively against highly scalable baselines.
\end{enumerate}

\subsubsection*{Simulation Procedure and Performance Metrics}
For each specific configuration defined by the dimension, sample size, graph type, and contamination scheme, we strictly follow this standardized procedure: First, we generate the true graph structure $\mathbf{K}^*$ using \texttt{huge.generator}. Second, for each of $R = 100$ independent Monte Carlo replicates, we:
\begin{enumerate}
    \item[(a)] generate $n$ i.i.d.\ observations from $\mathcal{N}(\bm{0}, (\mathbf{K}^*)^{-1})$,
    \item[(b)] apply the specified contamination scheme (if any) to 10\% of randomly selected observations,
    \item[(c)] fit the applicable competing methods (subject to a strict 60-second computational limit per replicate),
    \item[(d)] extract the estimated edge set $\hat{\mathcal{E}}$, defined as pairs $(i,j)$ where $|\hat{K}_{ij}| > 10^{-6}$, and compute all evaluation metrics, and
    \item[(e)] record the exact computational execution time.
\end{enumerate}
Finally, we compute the mean and standard error of the evaluation metrics across the $R=100$ replicates.

Graph recovery performance is evaluated using metrics that capture different aspects of estimation accuracy. Let TP, FP, FN, and TN denote true positives, false positives, false negatives, and true negatives for edge detection. To evaluate the support recovery performance, we report Precision, Recall~\citep{lingjaerde2024scalable}, $F_1$-score~\citep{ZhengAllen2024}, and the Matthews correlation coefficient (MCC)~\citep{ChenRenZhaoZhou2016,chicco2020advantages}, defined as follows:
\[
Precision=\frac{\text{TP}}{\text{TP} + \text{FP}},  \quad 
Recall = \frac{\text{TP}}{\text{TP} + \text{FN}}
\]
\[
F_1=\frac{2\text{TP}}{2\text{TP} + \text{FP} + \text{FN}}, \quad 
MCC = \frac{\text{TP} \cdot \text{TN} - \text{FP} \cdot \text{FN}}{\sqrt{({\text{TP}+\text{FP}})({\text{TP}+\text{FN}})({\text{TN}+\text{FP}})({\text{TN}+\text{FN}})}}
\]

Beyond statistical accuracy, the execution time recorded in step (e) allows us to assess algorithmic scalability. To ensure computational feasibility, we imposed a 60-second execution limit per replicate. Methods exceeding this threshold or failing to meet convergence criteria were recorded as non-convergent (NA). Consequently, alongside the estimation metrics, we report the Success Rate, defined as the proportion of Monte Carlo replicates that successfully yield a valid precision matrix estimate within the time constraint.

\subsection{Results for \texorpdfstring{$p \le 30$}{p <= 30}}
\label{subsec:results}

The extended simulation results, examining the performance of DROP across varying dimensions ($p \in \{10, 20, 30\}$) and sample sizes ($n \in \{250, 500, 1000\}$), are detailed in the Appendix. Across all examined dimensions, the relative performance rankings of the methods remain consistent. For brevity, we focus our discussion on the representative moderate-dimensional setting of $p=20$, highlighting the F1-score performance as a comprehensive measure of support recovery.

\begin{figure}[H]
    \centering
    \includegraphics[width=\textwidth]{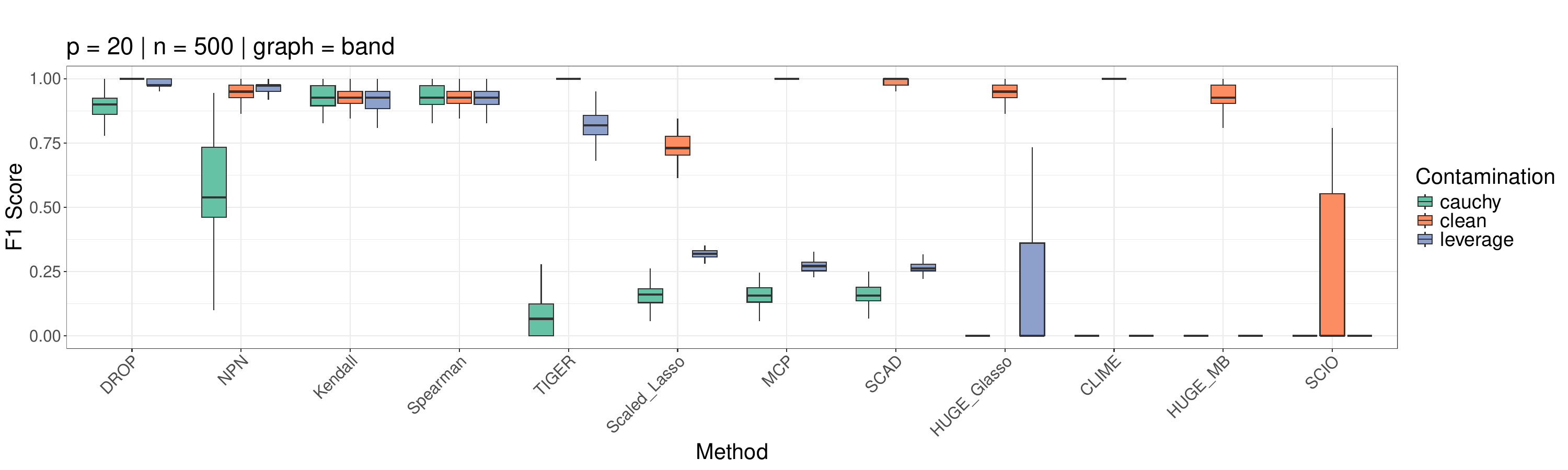}
    \caption{\footnotesize F1 Score comparison of the band graph with $p = 20$ nodes across contamination scenarios.}
    \label{fig:p20_box}
\end{figure}

Before delving into the numerical details of specific graph types, Figure \ref{fig:p20_box} provides a holistic visual summary of the estimation stability using the band graph as a representative example. The most striking observation across the three data regimes (Clean, Cauchy, and Leverage) is the consistently tight interquartile range of the DROP estimator. While competing robust methods exhibit significant performance fluctuations depending on the nature of the contamination, the estimates produced by DROP remain densely concentrated within a very narrow interval. This minimal estimation variance forms the core empirical advantage of our distributionally robust approach, demonstrating high reliability across diverse data generating mechanisms.

\subsubsection{Performance on Clean Data}
In the absence of data contamination, our primary goal is to verify that the robust formulation does not sacrifice statistical efficiency. Overall, DROP demonstrates highly competitive support recovery. When the data are generated from a true multivariate Gaussian distribution, DROP achieves high F1-scores that are generally on par with non-convex penalization methods (SCAD, MCP). For instance, at $n=500$, DROP yields excellent performance across Band, Hub, and Random graphs. We acknowledge that standard Glasso can yield slightly better support recovery in certain specific structures (e.g., the cluster graph at $p=30$, F1 = 0.72 vs. DROP's F1 = 0.66). However, the performance of DROP remains highly informative and comparable. This reflects a minor and acceptable efficiency-robustness trade-off, where the method sacrifices a marginal degree of accuracy under perfect Gaussianity to ensure stability against severe distributional perturbations.

\subsubsection{Robustness against Heavy-tailed Contamination}
The necessity of robust estimators becomes evident under Cauchy contamination, where standard methods (MB, Glasso, CLIME, SCIO) experience severe degradation, with F1-scores dropping to near-zero levels. 

In this regime, rank-based robust alternatives (such as NPN, Kendall, and Spearman) provide strong resilience against heavy-tailed noise. It is important to objectively acknowledge that in terms of the absolute mean F1-score, certain rank-based methods can outperform DROP under Cauchy contamination for specific graph types. However, as visually evidenced by the boxplots in Figure \ref{fig:p20_box}, DROP compensates for this by exhibiting significantly higher stability. While the performance of rank transformations can occasionally spread across a wider range across Monte Carlo replicates, DROP maintains a highly consistent, albeit sometimes slightly lower, recovery accuracy. DROP provides a more reliable and consistent estimation performance against extreme heavy-tailed outliers.

\subsubsection{Robustness against Leverage Point Contamination}
Leverage point contamination presents a unique challenge: the outliers preserve the underlying correlation structure but heavily distort the sample covariance matrix due to their extreme magnitude ($\mathcal{N}(\bm{0}, 100\bm{\Sigma})$). As shown in both the simulation tables and Figure \ref{fig:p20_box}, leverage points cause a noticeable decline in the performance of non-robust methods.

This is where DROP shows its greatest advantage. Although rank-based methods are also robust because they limit the impact of extreme values, DROP consistently matches or outperforms their mean F1-scores across various graph types. More importantly, DROP achieves this high accuracy while maintaining a very low estimation variance. This combination of high accuracy and low variance confirms that optimizing over a robust ambiguity set effectively protects the precision matrix estimation from outliers with inflated variances.

\subsubsection{Computational Efficiency}
All computational experiments were executed on a system equipped with an Apple M4 chip and 16 GB of unified memory. Evaluating the runtime for $p=20$ at $n=500$ under this environment, DROP requires an average of approximately 4 seconds. In contrast, the standard Glasso and the robust rank-based transformations (NPN, Kendall, Spearman) require approximately 10 seconds. These results suggest that the proposed algorithm achieves a favourable balance between computational efficiency and estimation stability.

\subsection{Performance on Large-scale Graphs \texorpdfstring{($p=100$ and $p=250$)}{(p=100 and p=250)}}
\label{subsec:large_scale_results}

\begin{figure}[p]
    \centering
    \begin{adjustbox}{max totalsize={\textwidth}{0.9\textheight},center}
        \begin{tabular}{cc}
        \includegraphics[width=0.43\textwidth]{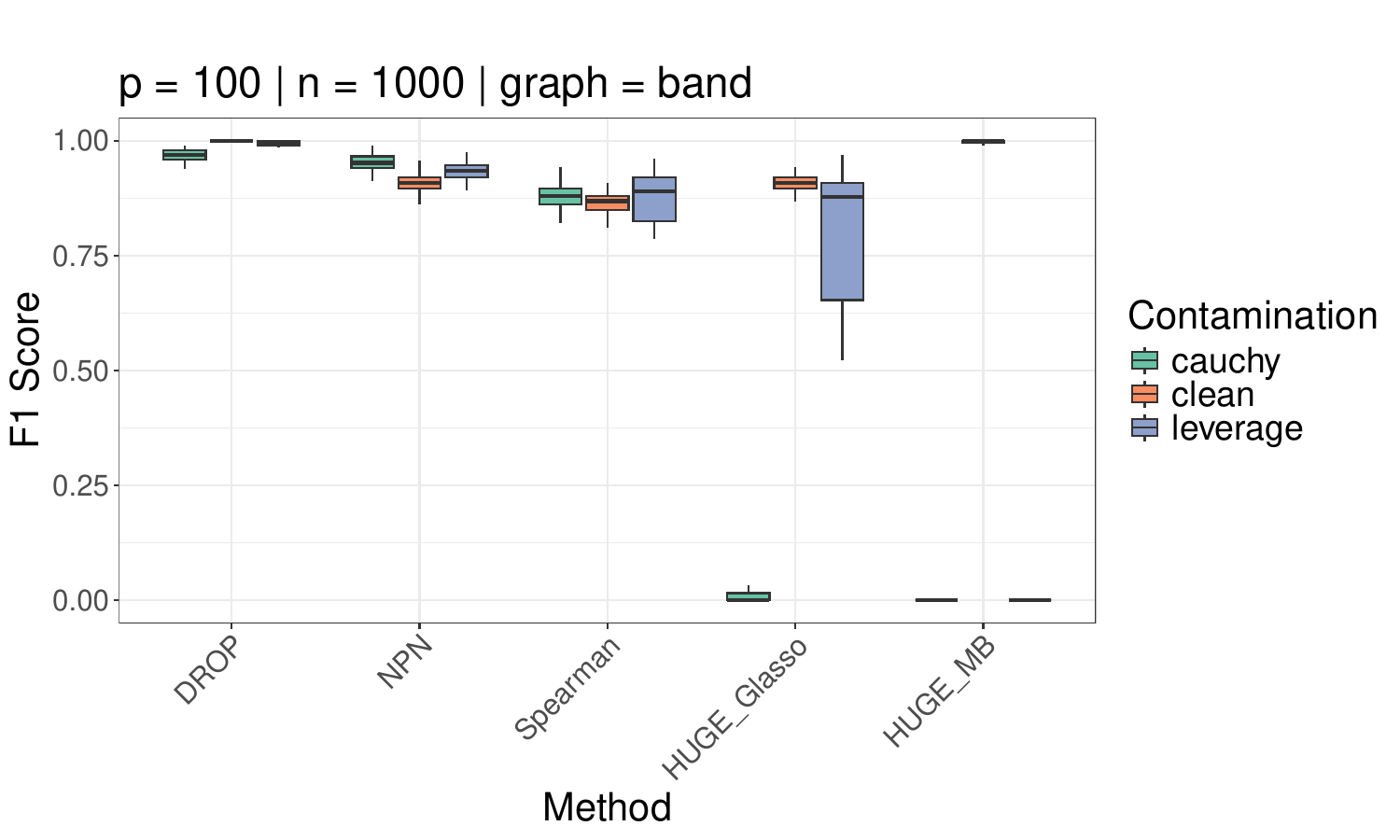} &
        \includegraphics[width=0.43\textwidth]{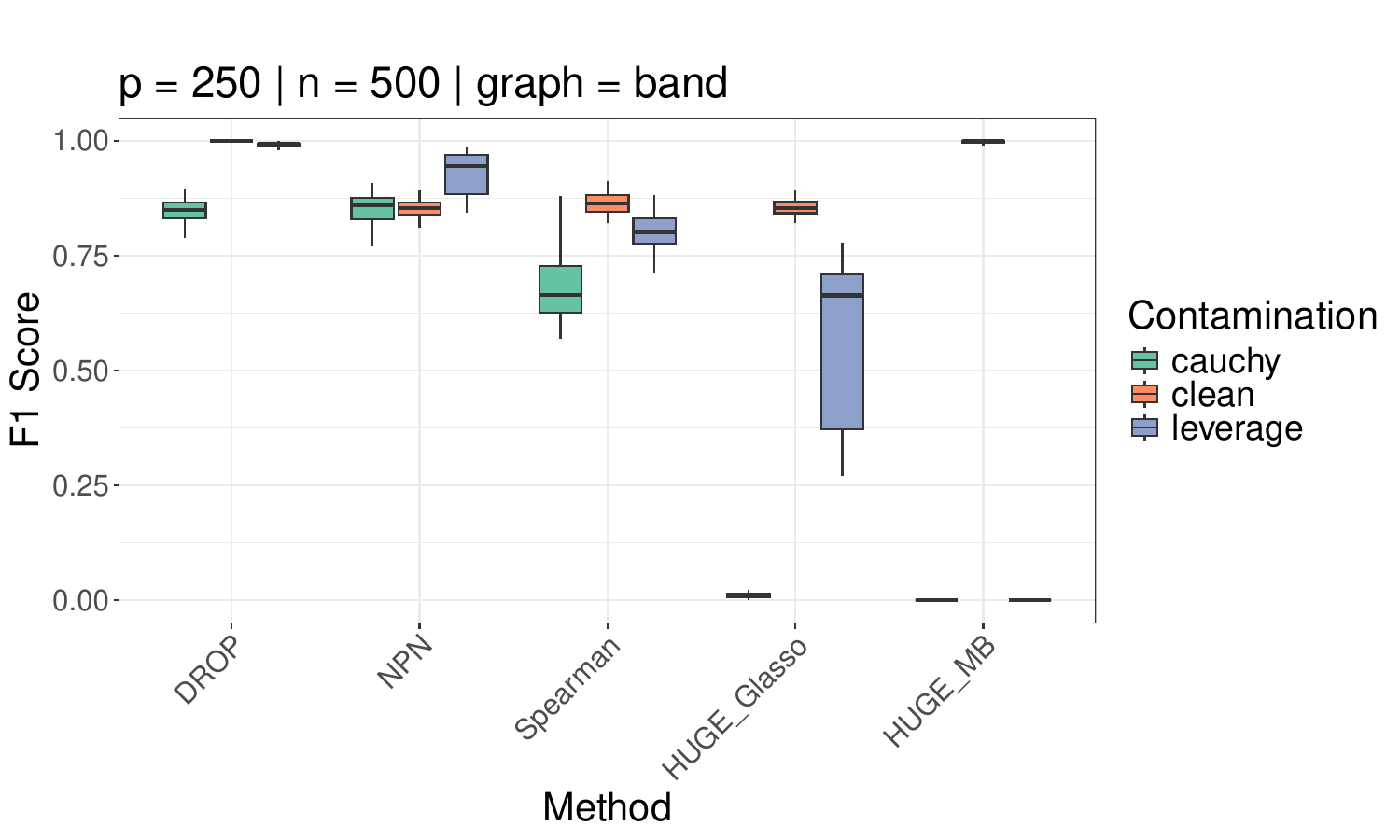} \\
        \includegraphics[width=0.43\textwidth]{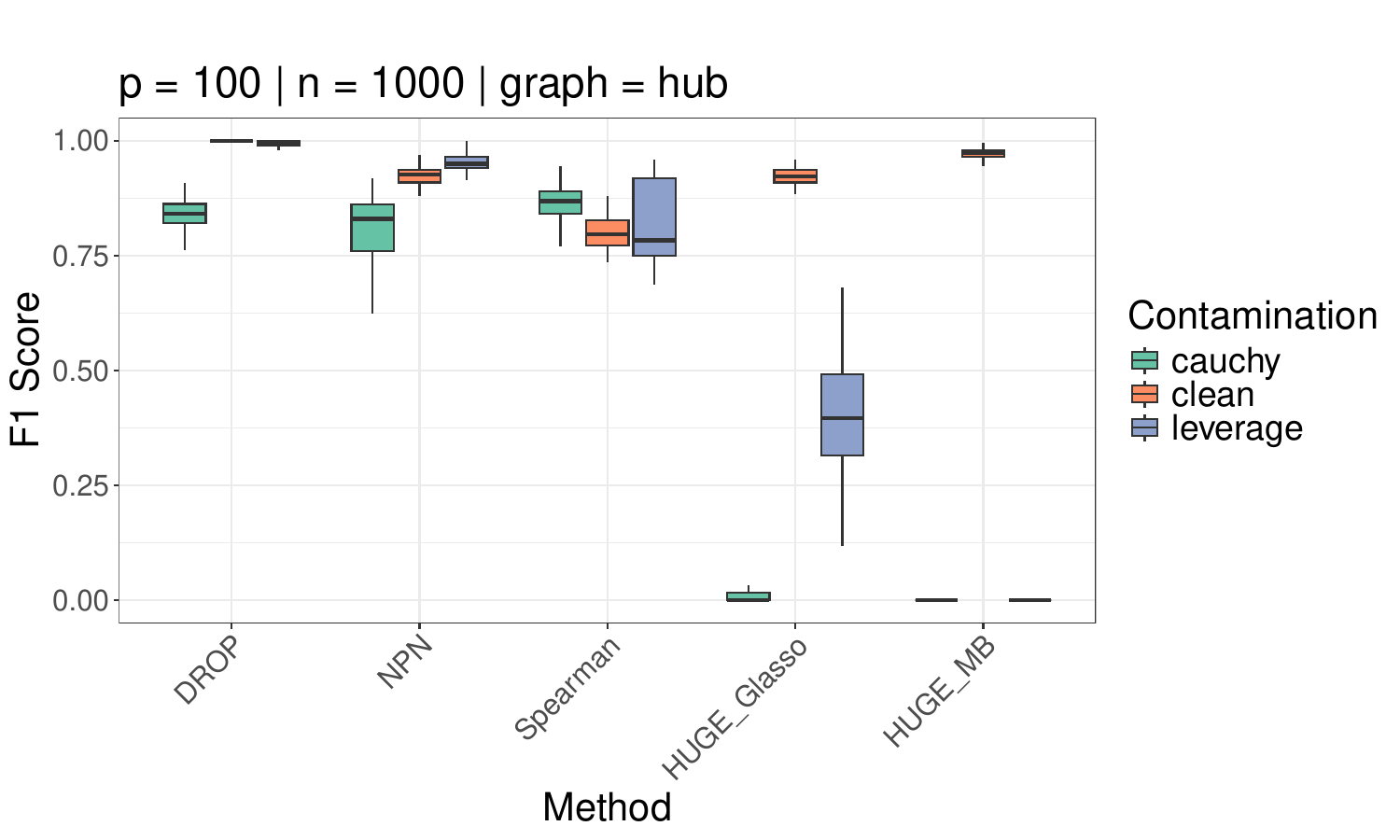} &
        \includegraphics[width=0.43\textwidth]{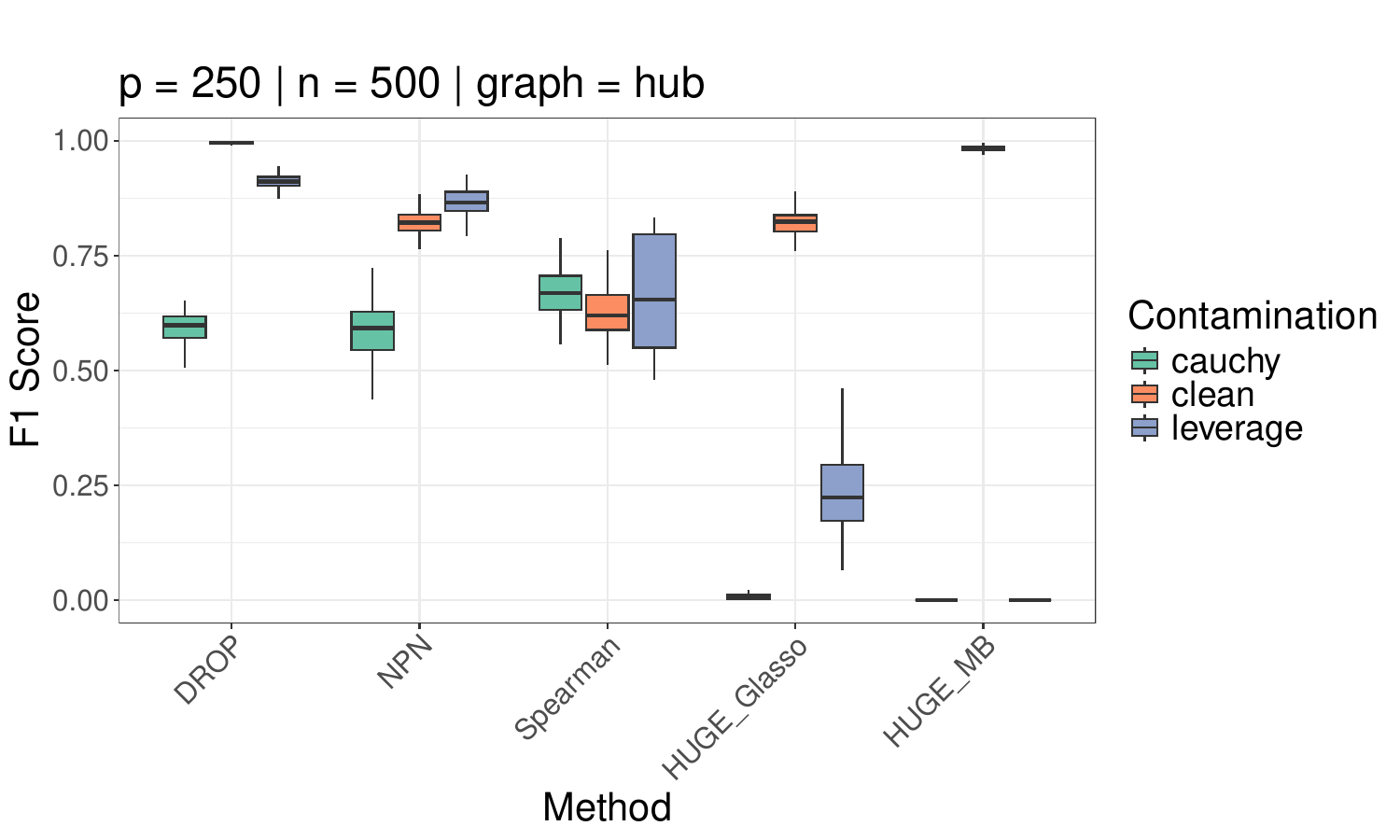} \\
        \includegraphics[width=0.43\textwidth]{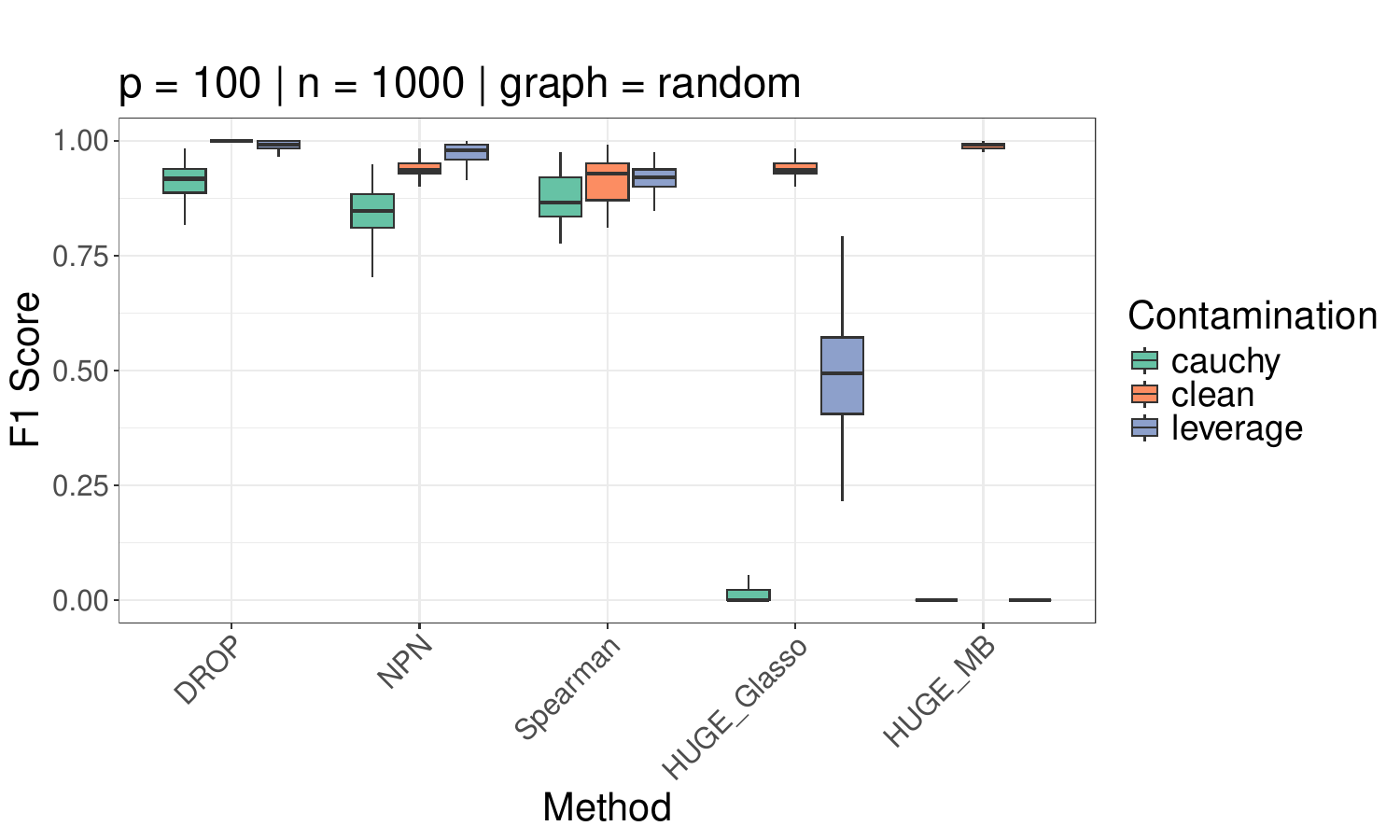} &
        \includegraphics[width=0.43\textwidth]{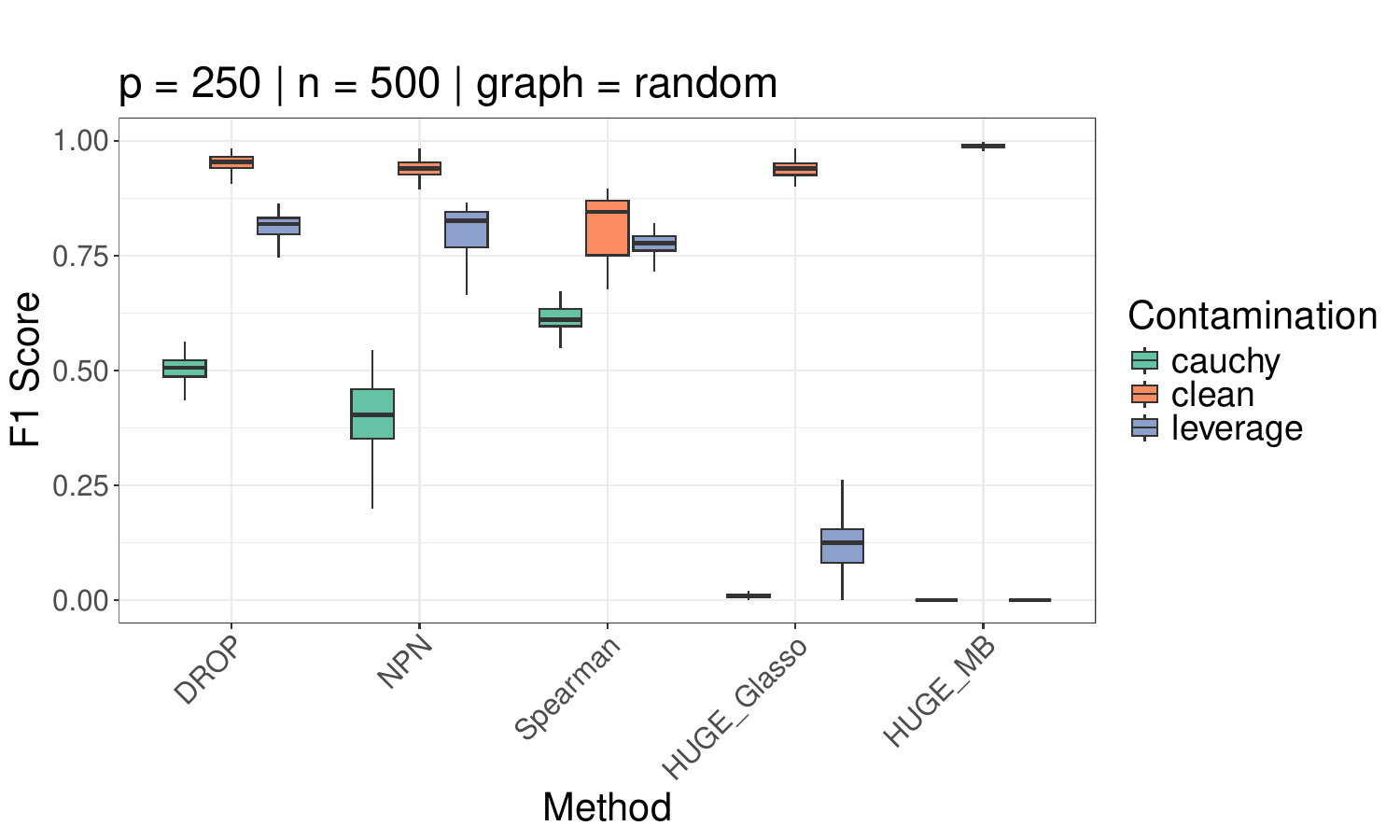} \\
        \includegraphics[width=0.43\textwidth]{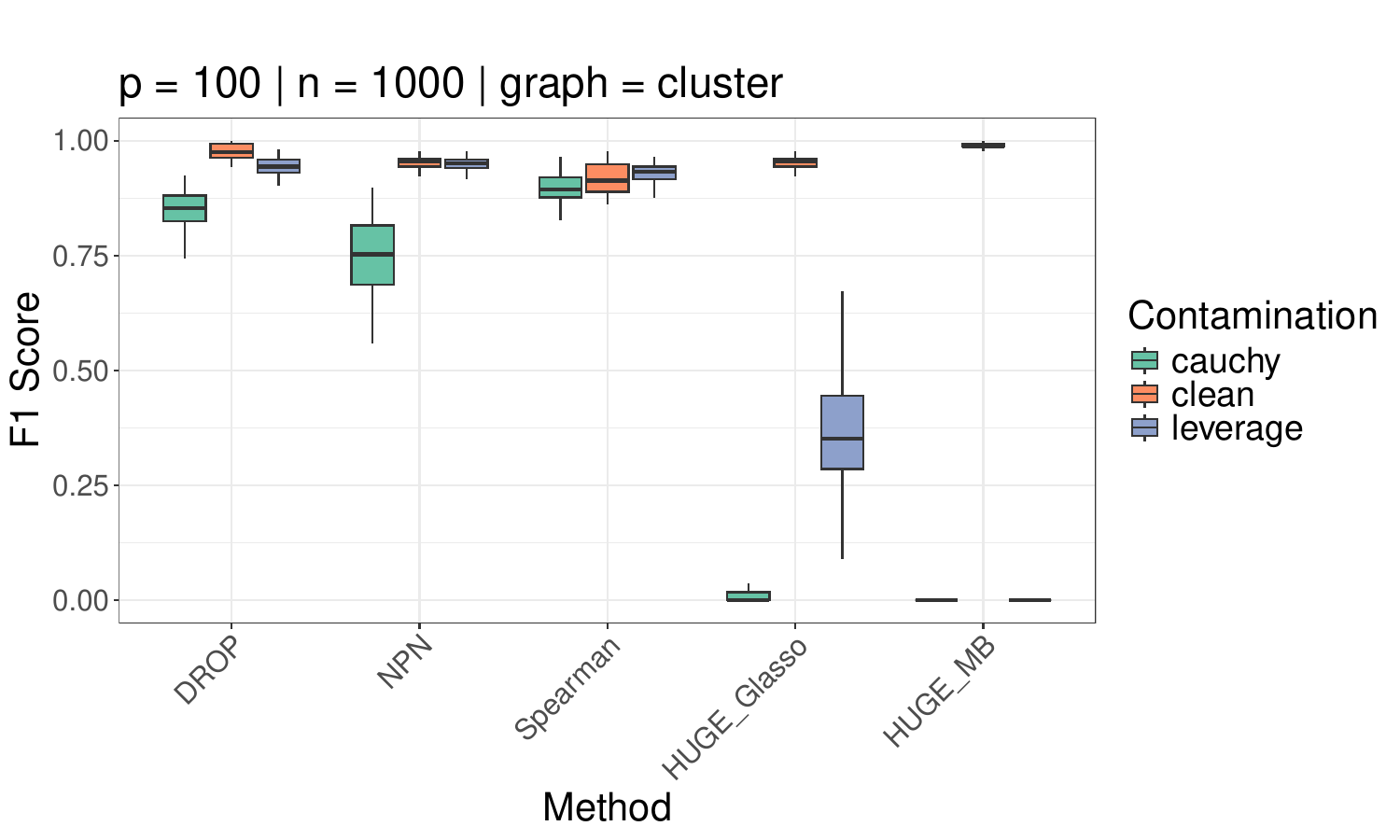} &
        \includegraphics[width=0.43\textwidth]{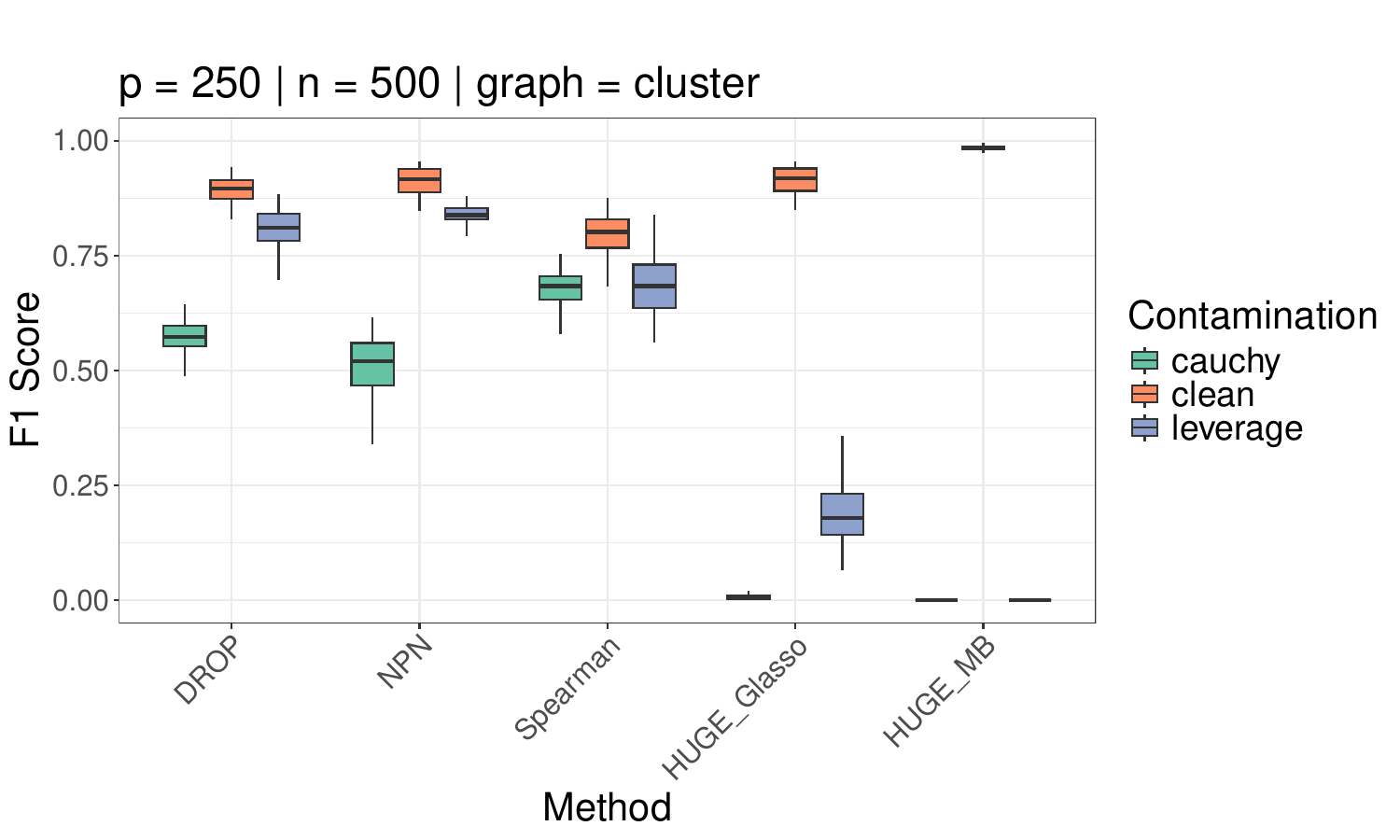} \\
        \includegraphics[width=0.43\textwidth]{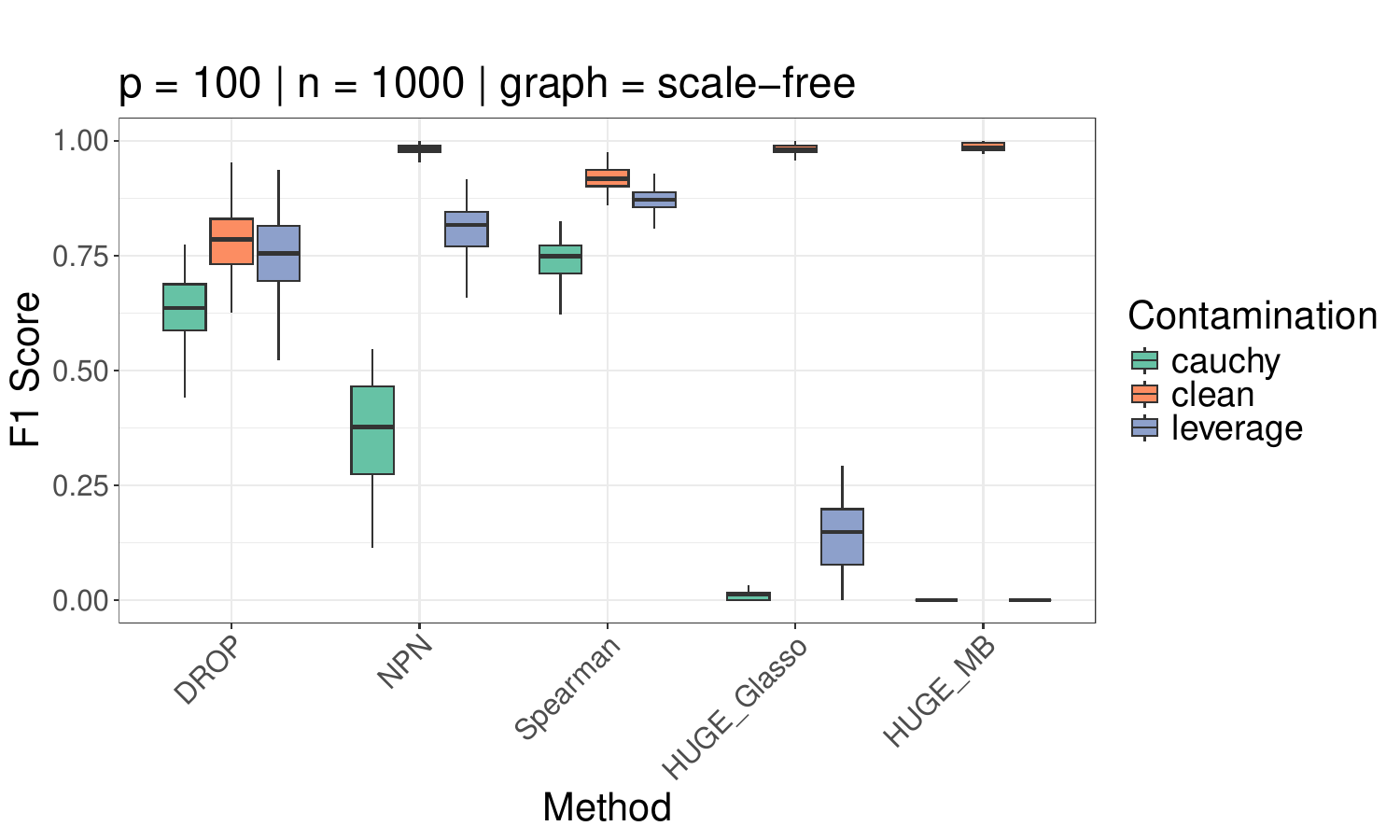} &
        \includegraphics[width=0.43\textwidth]{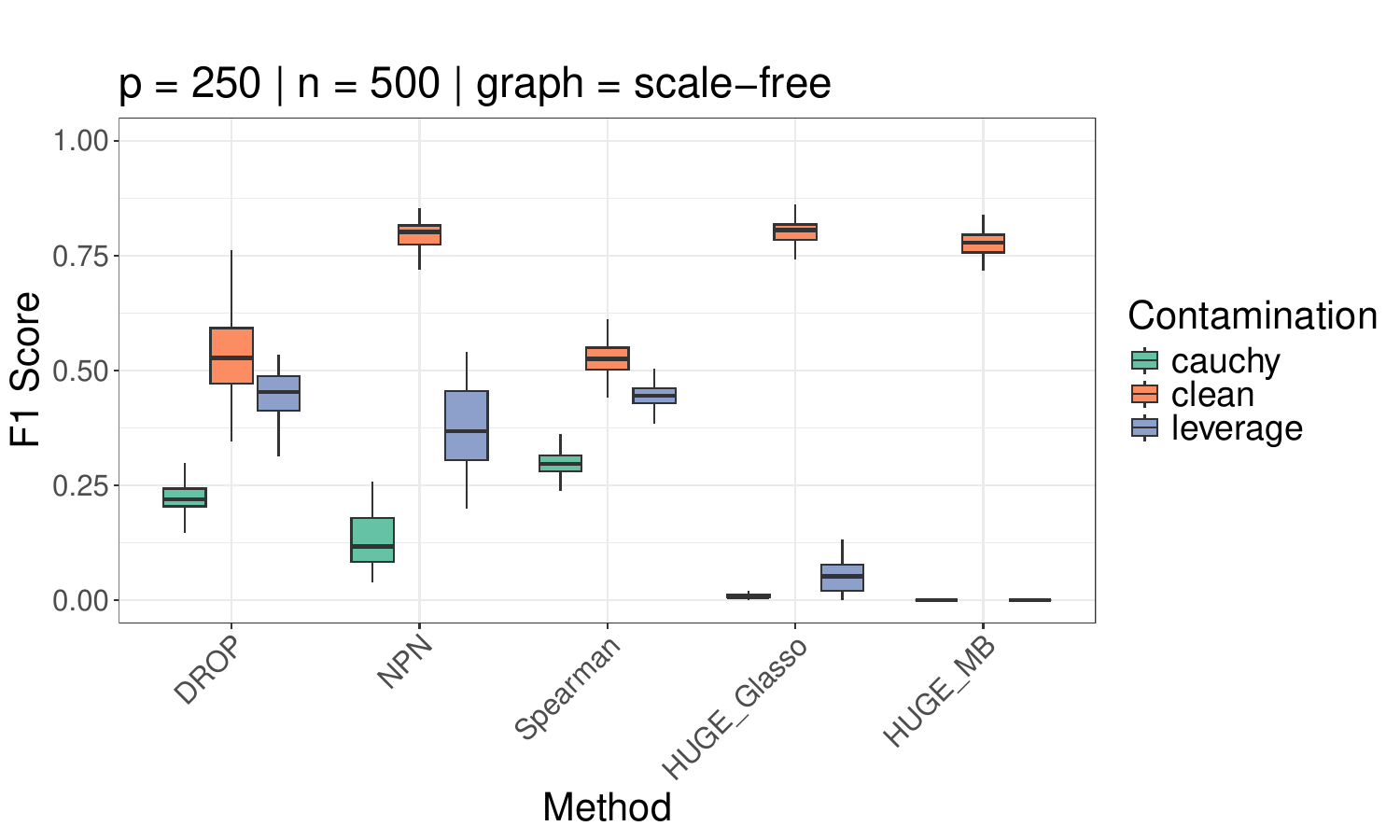}
        \end{tabular}
    \end{adjustbox}
    \caption{F1 score comparison across five graph structures. Left: $p=100$ ($n=1000$). Right: $p=250$ ($n=500$).}
    \label{fig:large_scale_all}
\end{figure}

To evaluate the scalability of DROP and to bridge the simulation studies with our subsequent real data application, we extend our evaluation to larger graph structures. Specifically, we consider scenarios with $p=100$ under sample sizes $n \in \{100, 1000\}$, and $p=250$ under $n \in \{500, 5000\}$. Due to the prohibitive computational cost in high-dimensional settings, methods that consistently exceeded the 60-second execution limit per replicate were omitted. Consequently, our comparison in this regime focuses on DROP alongside four scalable baselines: Glasso, MB, NPN, and Spearman. The support recovery performances for the representative cases of $p=100$ ($n=1000$) and $p=250$ ($n=500$) are visualized comprehensively in Figure \ref{fig:large_scale_all}. Detailed numerical results for all parameter configurations are provided in Appendix.

\subsubsection{High-dimensional Support Recovery and Robustness}

Estimating the graph structure is inherently more difficult in high-dimensional settings ($p=100$ and $p=250$). On clean data, DROP performs well in this regime. For standard graph structures, it achieves support recovery that matches non-robust baselines. Notably, in the specific setting of $p=250$ and $n=500$, DROP maintains high accuracy. This is evident in the hub graph, where DROP achieves an F1-score near 1.00. While the robust penalty induces some over-shrinkage in highly irregular structures (e.g., in the scale-free graph, Glasso achieves an F1 of 0.79 compared to DROP's 0.53), DROP still yields structurally informative estimates, confirming its applicability in high-dimensional settings.

When data contain outliers, non-robust methods (Glasso and MB) show a noticeable decrease in F1-scores across all structures (Figure \ref{fig:large_scale_all}). This outcome is expected, as these estimators are not designed to handle heavy-tailed distributions or variance inflation.

For $p=100$, DROP demonstrates strong performance across the band, hub, random, and cluster graphs, particularly under leverage point contamination. Under Cauchy noise, while Spearman achieves higher mean F1-scores in the hub and cluster graphs, DROP remains highly comparable and maintains a relatively smaller estimation variance. In the scale-free graph, Spearman shows better recovery; however, DROP continues to provide a reliable baseline without extreme performance degradation.

As the dimension increases to $p=250$, the relative advantages of the methods vary with the graph structure. Under leverage contamination, DROP performs well on the band, hub, and random graphs. In the scale-free and cluster graphs under leverage noise, the performance of DROP is comparable to Spearman and NPN, respectively. Notably, Spearman yields lower F1-scores on the hub and cluster graphs in this setting, indicating that its performance varies with the underlying graph structure. Under Cauchy noise at $p=250$, DROP performs well on the band graph, though Spearman outperforms it on the remaining structures.

These detailed comparisons highlight the core advantage of DROP: its overall stability and robustness. While rank-based methods such as NPN and Spearman may excel in specific combinations of graph structures and contamination types, their performance and estimation variance fluctuate substantially across scenarios. DROP, conversely, avoids these severe fluctuations. By maintaining consistent structural recovery and stable behaviour across diverse conditions, DROP provides a dependable approach for robust Gaussian graphical model estimation. This combination of stability and robustness supports its application to real-world data, where the exact nature of the structural noise is typically unknown.

\subsubsection{Computational Scalability}
For $p=250$ and $n=500$, DROP requires an average of 18 to 44 seconds to converge across the evaluated graph structures. This runtime is higher than that of Glasso, NPN, and Spearman (11 to 26 seconds), but remains within the 60-second limit. The additional execution time is a standard trade-off for the improved stability under data contamination, making the algorithm suitable for higher-dimensional datasets.

\section{Real Data Application: Functional MRI Connectivity}
\label{sec:real_data}

Motivated by the strong empirical performance and scalability of DROP in the high-dimensional simulations (particularly the $p=250$ setting), we apply our proposed method to a real functional magnetic resonance imaging (fMRI) dataset. In neuroimaging studies, fMRI recordings are frequently confounded by physiological noise, scanner artifacts, and subject head motion, often resulting in heavy-tailed noise distributions and signal outliers. Consequently, estimating functional brain connectivity networks presents a high-dimensional challenge with inherent distributional uncertainty, making it an ideal testbed for the DROP estimator.

\subsection{Dataset and Problem Setup}
We utilize the publicly available Music Genre fMRI Dataset~\citep{ds003720}, which captures the Blood Oxygenation Level Dependent (BOLD) signals of subjects listening to diverse musical stimuli. To construct the region-wise connectivity network, the brain was parcellated into $p=264$ regions of interest (ROIs) based on the standard Power-264 atlas~\citep{Power2010}.

\begin{figure}[htbp]
    \centering
    \includegraphics[width=\textwidth]{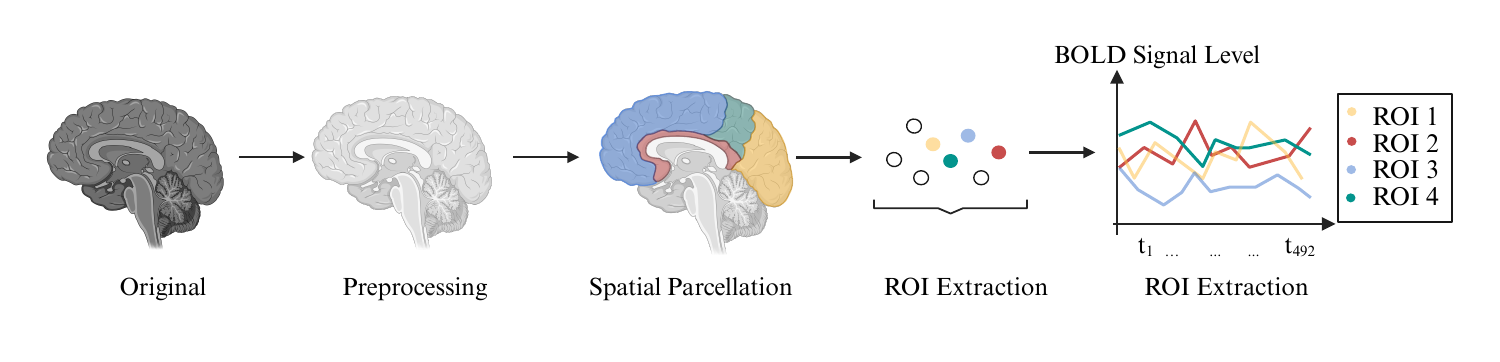}    
    \caption{\footnotesize Workflow of fMRI data extraction from raw images to region-wise BOLD time series based on the Power-264 parcellation (created in  https://BioRender.com)}
    \label{fig:processing}
\end{figure}

Figure~\ref{fig:processing} provides an overview of the fMRI data extraction workflow, from raw images to region-wise BOLD time series using the Power-264 parcellation. The process begins with standard fMRI preprocessing steps, including slice-timing correction to adjust for inter-slice acquisition delays, motion correction via realignment to the mean functional image, coregistration to structural images, normalization to MNI space, and spatial smoothing with a Gaussian kernel to improve signal-to-noise ratio. Following preprocessing, the data undergo spatial parcellation into 264 regions of interest (ROIs). For each ROI, the mean BOLD signal is extracted across time, yielding region-wise time series that capture the temporal dynamics of neural activity. As illustrated in the figure, this results in a structured output where each ROI corresponds to a column vector of BOLD signal intensity across sequential 492 time points, forming the basis for subsequent connectivity analyses.

Since real fMRI data lack a ground-truth graph, we assess the robustness of the estimators by directly applying contamination to the empirical dataset. First, we estimate the baseline connectivity networks using the original, clean fMRI data. To evaluate performance under non-ideal conditions, we then corrupt the fMRI design matrix with the same Cauchy and leverage point contamination ($r = 0.1$) utilized in Section~\ref{sec:simulations}. We compare DROP against the standard GLASSO, MB, NPN, and Spearman estimators.

\subsection{Sparse Connectivity and Modular Organization}
\label{subsec:modularity}

\begin{figure}[htbp]
    \centering
    \includegraphics[width=0.9\textwidth]{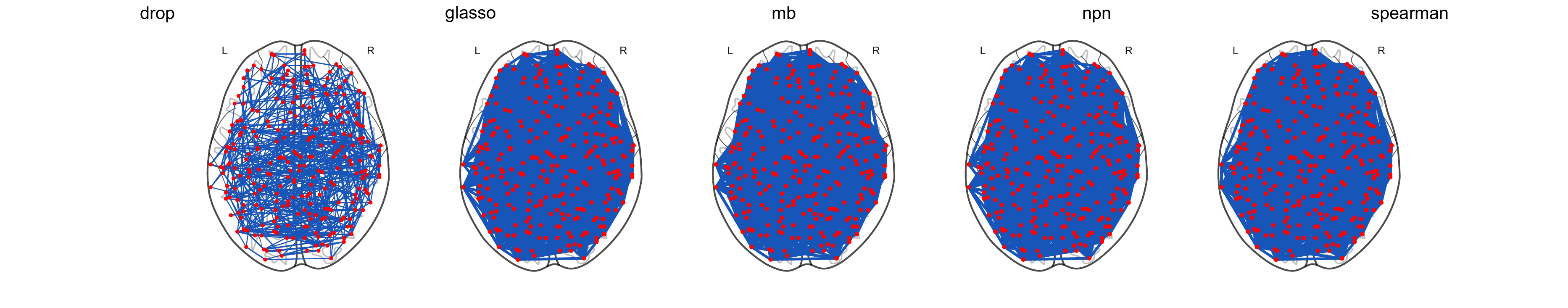}\\
    {\footnotesize (a) sub-001}\\[0.5em]

    \includegraphics[width=0.9\textwidth]{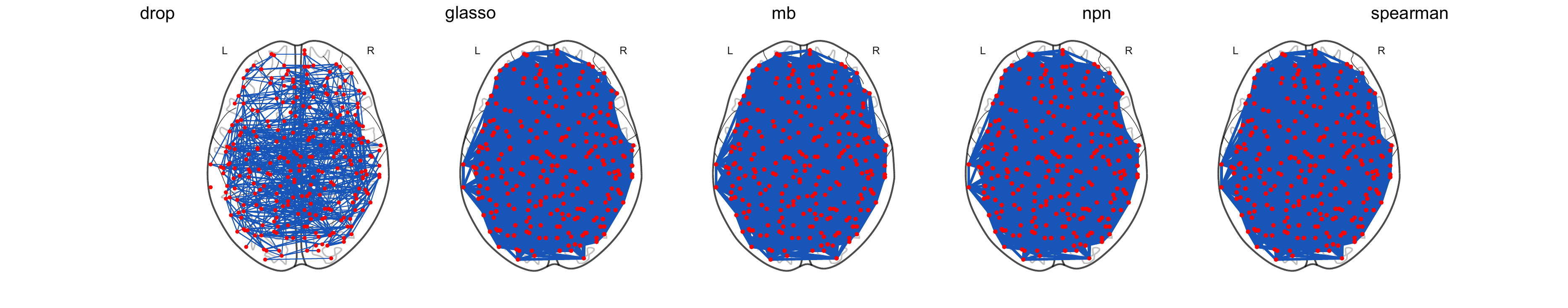}\\
    {\footnotesize (b) sub-002}\\[0.5em]

    \includegraphics[width=0.9\textwidth]{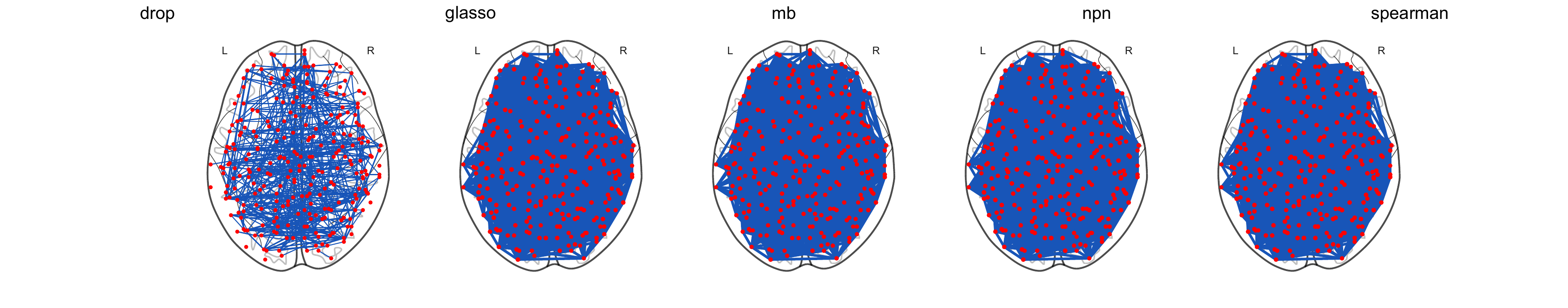}\\
    {\footnotesize (c) sub-003}

    \caption{\footnotesize Estimated brain connectivity networks for three subjects under the clean data scenario. Auditory regions are highlighted in blue.}
    \label{fig:brain_connection}
\end{figure}

\begin{figure}[htbp]
    \centering
    \includegraphics[width=0.6\textwidth]{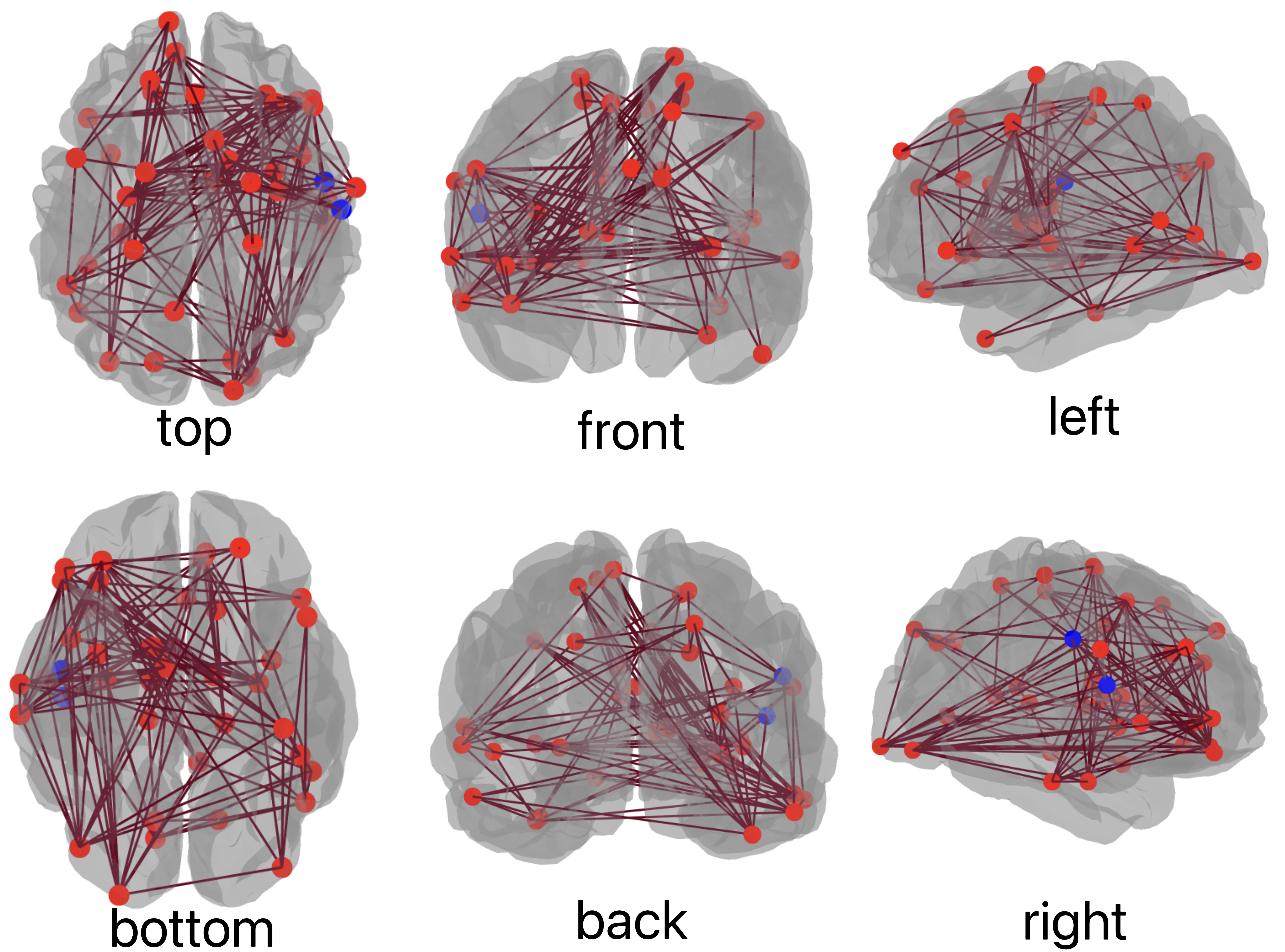}    
    \caption{\footnotesize 3D spatial visualization of the strongest 20 ROI and connectivity network estimated by the proposed DROP method.}
    \label{fig:3d}
\end{figure}

To investigate the structural properties of the estimated networks, we visualize the brain connectivity matrices produced by DROP, GLASSO, MB, NPN, and Spearman. Figure~\ref{fig:brain_connection} displays the estimated connectomes for three subjects under the clean data condition. Following the thresholding practice in neuroimaging literature~\citep{Zhang2019Aberrant}, only edges with an absolute partial correlation strength above 0.005 are retained to highlight the most significant functional connections. Nodes corresponding to auditory regions are colored in blue, while other regions are in red. To provide a clearer anatomical perspective, Figure~\ref{fig:3d} additionally displays a 3D spatial mapping of the network estimated specifically by the proposed DROP method.

As observed in Figure~\ref{fig:brain_connection}, DROP estimates a sparser network than the competing methods. Statistically, inducing such sparsity helps control false positive edges and prevent overfitting in high-dimensional regimes. This conservative structural recovery is consistent with evidence in network neuroscience, which indicates that functional brain networks are intrinsically sparse~\citep{Huang2010, Ryali2012}.

To quantitatively evaluate the graph structure, we compare the network modularity alongside the total number of edges. Modularity measures how well a network separates into densely connected communities with sparse connections between them. Formally, the modularity score $\mathcal{M}$ is defined as~\citep{Newman2006}:
\begin{equation}
\mathcal{M} = \frac{1}{2|\mathcal{E}|} \sum_{i,j=1}^p
\left(
A_{ij} - \frac{d_i d_j}{2|\mathcal{E}|}
\right)
\mathbb{I}(g_i = g_j),
\label{eq:modularity}
\end{equation}
where $A_{ij}$ is the adjacency matrix, $d_i$ and $d_j$ denote the degrees of nodes $i$ and $j$, $|\mathcal{E}|$ is the total number of edges in the estimated graph, and $\mathbb{I}(\cdot)$ is the indicator function such that $\mathbb{I}(g_i = g_j) = 1$ if nodes $i$ and $j$ belong to the same community, and $0$ otherwise, with $g_i$ representing the community assignment of node $i$ (identified via the Louvain algorithm~\citep{Blondel2008}) and 0 otherwise. Higher values of $\mathcal{M}$ indicate a more pronounced community structure.

\begin{figure}[htbp]
    \centering

    \includegraphics[width=0.95\textwidth]{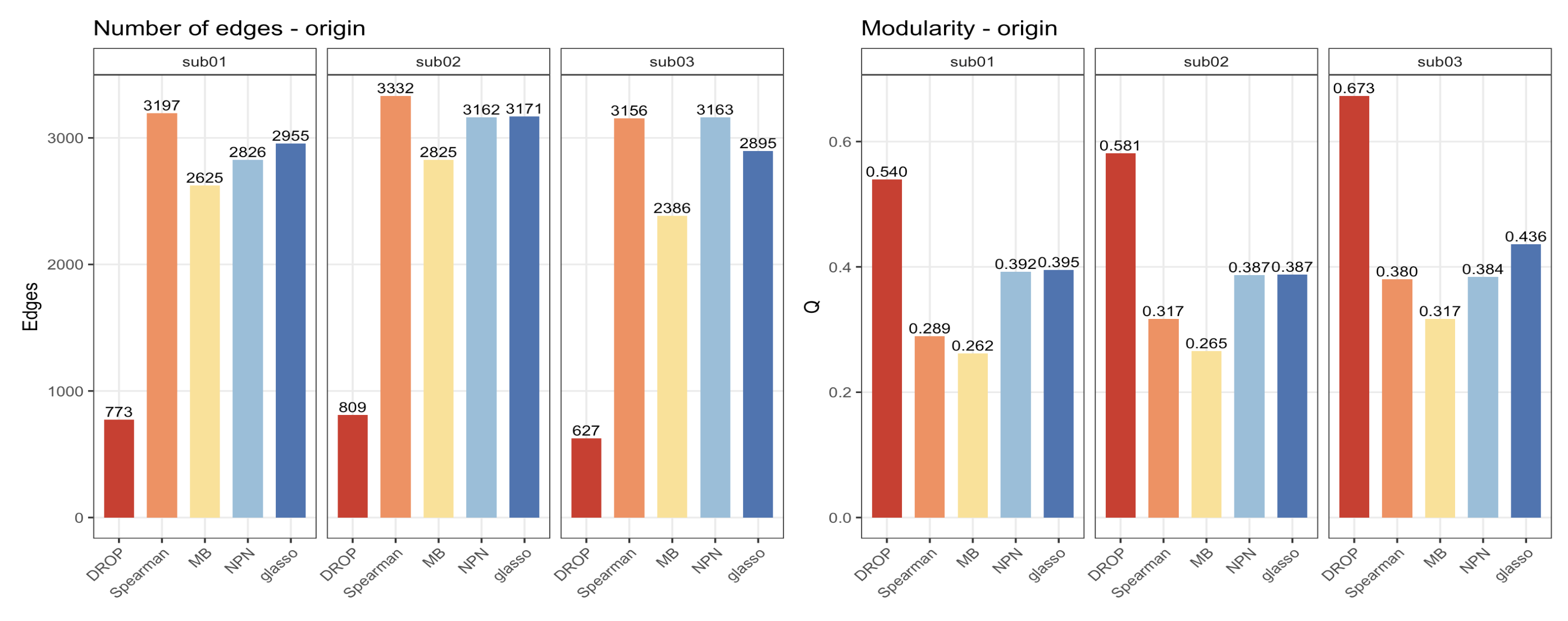}\\
    {\footnotesize (a) Clean (Origin)}\\[0.5em]

    \includegraphics[width=0.95\textwidth]{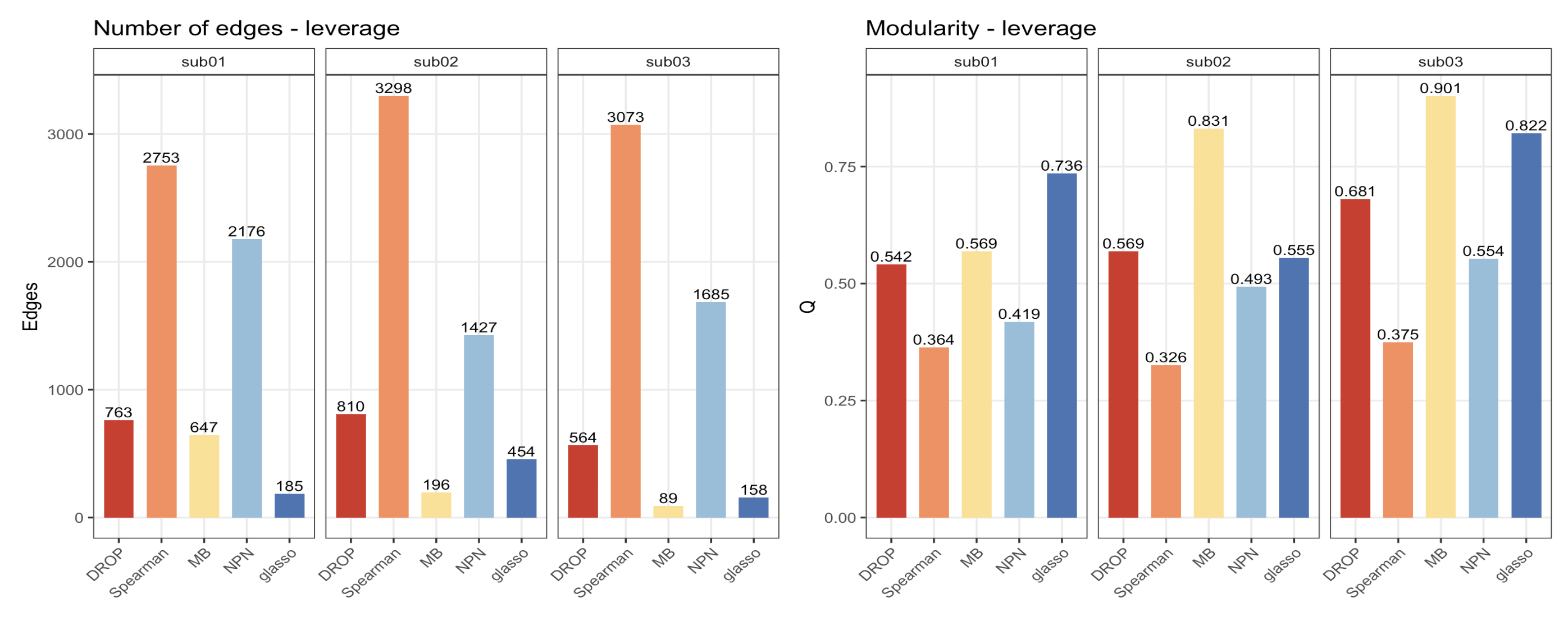}\\
    {\footnotesize (b) Leverage Contamination}\\[0.5em]

    \includegraphics[width=0.95\textwidth]{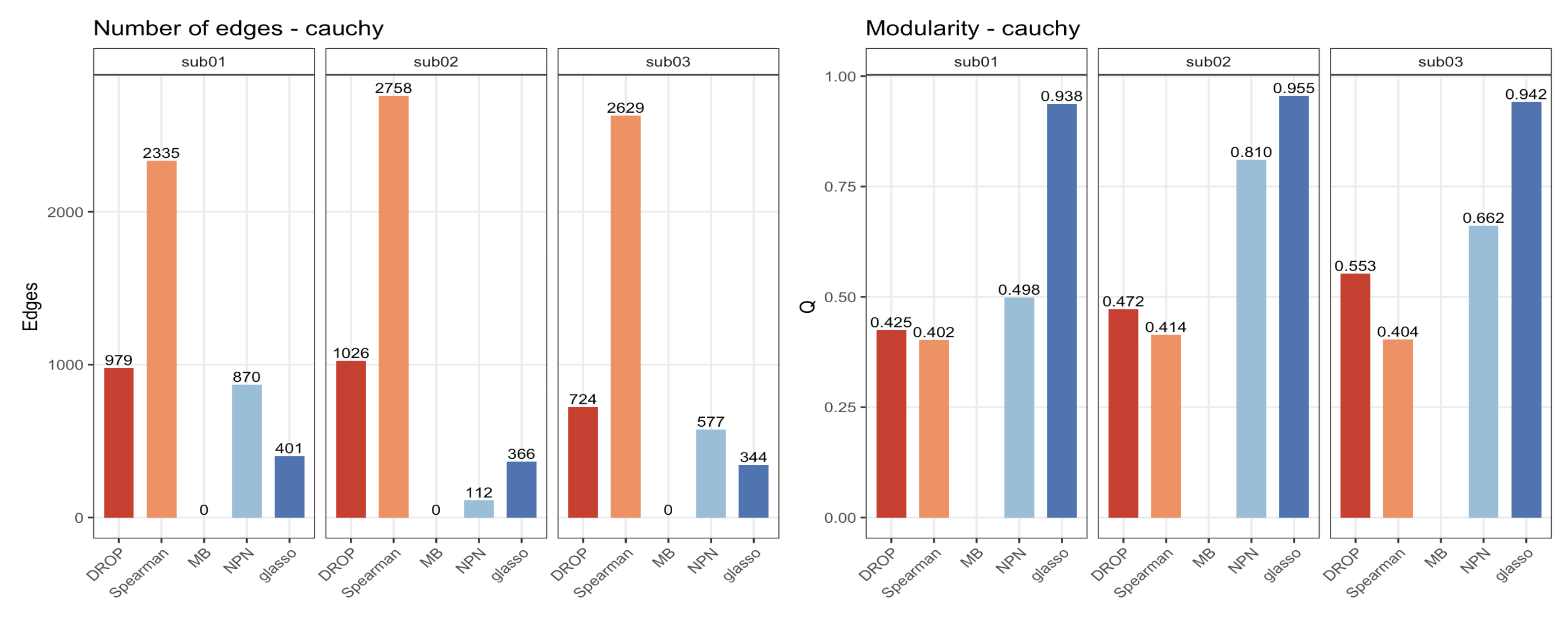}\\
    {\footnotesize (c) Cauchy Contamination}

    \caption{\footnotesize Comparison of the number of estimated edges (left axis) and modularity scores (right axis) across the five estimation methods under different contamination scenarios.}
    \label{fig:edges}
\end{figure}

Figure~\ref{fig:edges} summarizes the network sparsity and modularity scores across all data conditions. Under the clean setting, DROP consistently yields the sparsest networks alongside the highest modularity. For example, in Subject 001, DROP estimates fewer edges (773) than competing methods while achieving the highest modularity ($\mathcal{M} = 0.540$). This suggests that the edges retained by DROP form a coherent community structure rather than a dense, unorganized graph.

Under data contamination, DROP maintains stable performance. With leverage perturbation, its edge counts and modularity values remain similar to those obtained from the clean data. In contrast, non-robust estimators like GLASSO and MB show larger variations, reflecting their sensitivity to outliers. Under heavy-tailed Cauchy noise, while all methods experience performance degradation, DROP preserves a sparser graph and higher modularity than the baselines. The competing methods tend to produce denser networks with lower modularity scores, indicating the inclusion of noise-induced false positive edges.

\subsection{Node Degree Distribution Across Functional Systems}
\label{subsec:degree_distribution}

To examine the behavior of the estimators across localized brain regions, we analyze the node degree distributions grouped by the functional systems defined in the Power-264 atlas. For an estimated adjacency matrix $A$, the degree of node $i$ is defined as $d_i = \sum_{j \neq i} A_{ij}$, representing the local connectivity density. Given the experimental context of music genre recognition, the auditory network is of particular interest, alongside other major functional modules.

\begin{figure}[htbp]
    \centering
    \includegraphics[width=\textwidth]{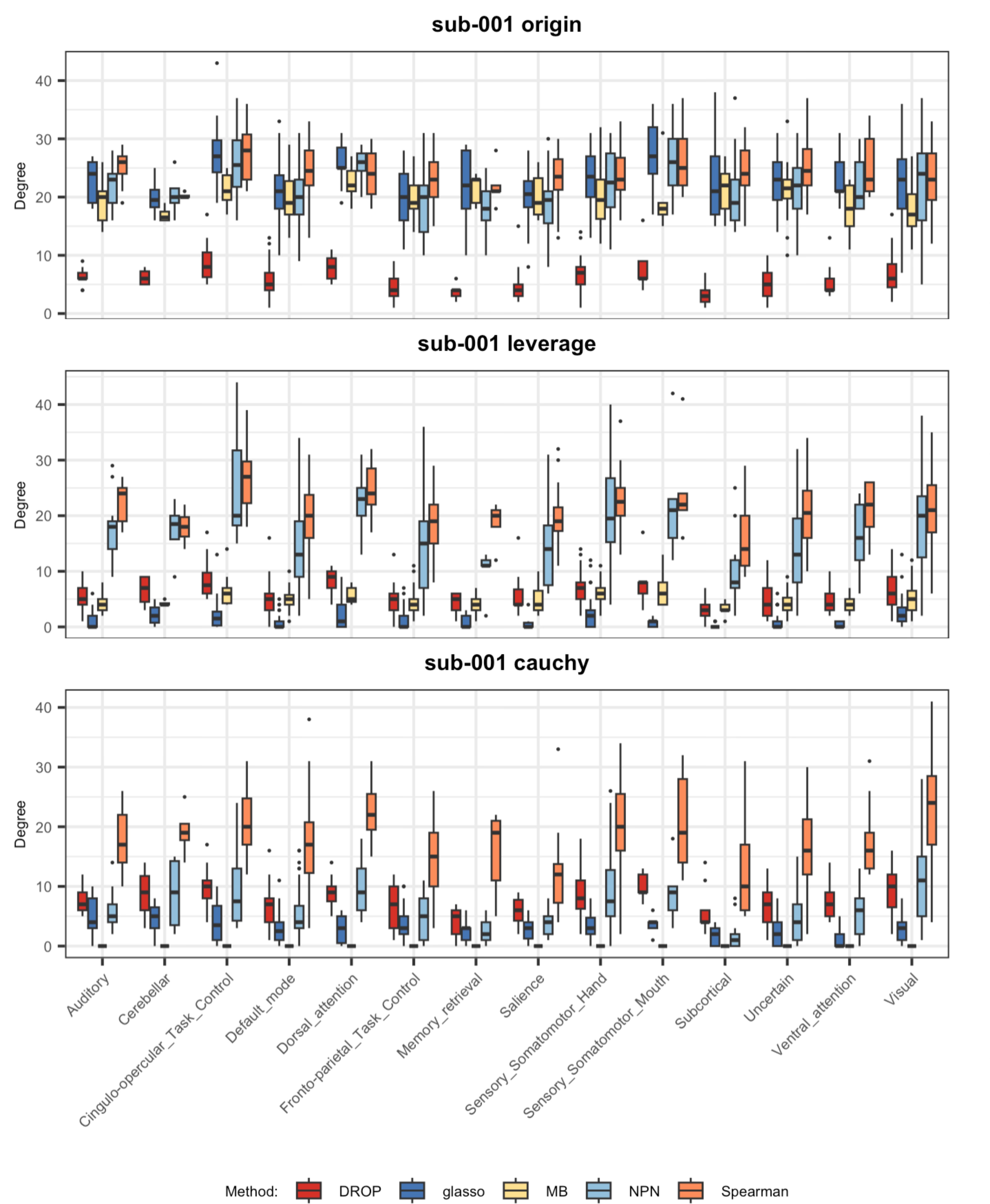}    
    \caption{\footnotesize Node degree distributions across functional systems for Subject 001 under Clean, Leverage, and Cauchy contamination.}
    \label{fig:sub01 box plot}
\end{figure}

Figure~\ref{fig:sub01 box plot} displays the degree distributions for Subject 001 under the clean, leverage, and Cauchy contamination scenarios. The corresponding box plots for the remaining subjects, along with the system-level mean degree comparisons across all conditions, exhibit similar patterns and are provided in Appendix.

The degree distributions show that DROP yields the lowest median degrees and the narrowest interquartile ranges across all functional systems. This sparsity confirms a more conservative criterion for edge selection compared to Spearman-based networks, which produce the densest graphs. Standard GLASSO, NPN, and MB estimators generally fall between these two extremes. Under data contamination, the degree profiles of non-robust methods show noticeable shifts. Specifically, leverage points cause an upward shift in the degrees for several methods, while Cauchy noise increases the dispersion of the estimates, as evidenced by elongated whiskers and outliers in Figure~\ref{fig:sub01 box plot}. In contrast, the distributions for DROP remain stable across both perturbation schemes, reflecting its resistance to variance inflation and heavy-tailed noise.

The results also capture heterogeneity across brain systems. Integrative and sensory networks, such as the Default Mode and Visual systems, exhibit higher node degrees across most methods, suggesting these regions act as connectivity hubs. The Auditory system shows lower median degrees and localized connectivity. Notably, the estimation in Subcortical and Uncertain regions appears more sensitive to data quality and the choice of regularization, as shown by the increased variability under contamination. Overall, the fMRI application is consistent with the simulation results in Section~\ref{sec:simulations}. Under clean conditions, DROP identifies a sparse network with modular organization, and under heavy-tailed or leverage-point contamination, it maintains structural stability while competing methods are prone to noise-induced false positive edges. Taken together, the reliable network structures recovered by DROP suggest that music perception is supported by coordinated interactions across distributed functional systems, rather than isolated processing within the auditory cortex alone.


\section{Discussion}
\label{sec:discussion}

In this paper, we developed DROP, a robust methodology for estimating high-dimensional Gaussian Graphical Models. Motivated by the vulnerability of standard precision matrix estimators, DROP is designed to enforce structural sparsity while maintaining robustness against general data contamination, such as extreme outliers and heavy-tailed noise. Theoretically, we established error bounds for the proposed estimator. To facilitate practical application and reproducible research, the proposed methodology is implemented in the R package \texttt{DROP}, which will be made publicly available on GitHub.

The empirical performance of DROP was evaluated through both high-dimensional simulations and a real-world fMRI application. In our simulations, DROP demonstrated an ability to control the rate of false positive edges. When the data are subjected to leverage points or heavy-tailed Cauchy noise, conventional estimators such as GLASSO and MB tend to produce denser networks with reduced modularity, reflecting their sensitivity to such perturbations. In contrast, DROP identifies a more stable graph structure, maintaining consistent sparsity and localized degree distributions. This reliability is beneficial for neuroimaging analyses, where scanner artifacts and physiological noise often deviate from standard Gaussian assumptions.

Several directions remain open for future research. While our current implementation focuses on standard precision matrix estimation via nodewise regression, the underlying distributionally robust optimization formulation naturally extends to a broader class of graphical models. For instance, this robust approach can be adapted to models with structural symmetries, such as colored Gaussian graphical models~\citep{hojsgaard2008graphical,roverato2024exploration,li2025Constraints}, which impose equality constraints on the precision matrix. Furthermore, incorporating the DROP framework into Gaussian graphical regressions could enable the robust estimation of covariate-dependent precision matrices~\citep{zhang2025multi}. This would provide a reliable tool for estimating heterogeneous networks whose edge weights are modulated by high-dimensional external covariates, even when the observed data are contaminated.

\section*{Data Availability and Reproducibility}

The fMRI dataset used in this study is publicly available from the OpenNeuro repository under accession number \texttt{ds003720}, version 1.0.1 (\url{https://openneuro.org/datasets/ds003720/versions/1.0.1}). The analysis focuses on three subjects (sub-001, sub-002, and sub-003), specifically the functional runs corresponding to the \texttt{task-Training} phase. To reproduce the results, users should download the original BIDS-formatted dataset directly from OpenNeuro. All preprocessing pipelines, ROI time series extraction scripts, and the R code for the DROP graphical model estimation are provided in the supplementary materials (and our public GitHub repository).

\acks{The authors would like to thank Mengyu Xie for valuable assistance with the literature review and Yiting Chen for helpful comments on the manuscript. Qiong Li’s Research partially supported by NSFC (Grant No.
12271047) and Guangdong and Hong Kong Universities “1+1+1” Joint Research Collaboration
Scheme (2025A0505000010) and Guangdong provincial
key laboratory of interdisciplinary research and application for data
science (2022B1212010006). Wenzhi Yang was supported by the National Natural Science Foundation of China (NSFC, Grant No.~12471248). Xiaoping Shi was supported by the Natural Sciences and Engineering Research Council of Canada (NSERC) Discovery Grant (RGPIN-2022-03264), the NSERC Alliance International Catalyst Grant (ALLRP-590341-23), and the University of British Columbia Okanagan (UBC-O) Vice-Principal Research in collaboration with the Irving K.~Barber Faculty of Science.}

\bibliography{refs}

@article{AvellaMedina2018,
  author  = {Avella-Medina, Marco and Battey, Heather and Fan, Jianqing and Li, Qiwei},
  title   = {Robust estimation of high-dimensional covariance and precision matrices},
  journal = {Biometrika},
  volume  = {105},
  number  = {2},
  pages   = {271--284},
  year    = {2018}
}

@article{banerjee2008model,
  title={Model selection through sparse maximum likelihood estimation for multivariate Gaussian or binary data},
  author={Banerjee, Onureena and El Ghaoui, Laurent and d'Aspremont, Alexandre},
  journal={The Journal of Machine Learning Research},
  volume={9},
  pages={485--516},
  year={2008},
  publisher={JMLR. org}
}

@article{barabasi1999emergence,
  title={Emergence of scaling in random networks},
  author={Barab{\'a}si, Albert-L{\'a}szl{\'o} and Albert, R{\'e}ka},
  journal={Science},
  volume={286},
  number={5439},
  pages={509--512},
  year={1999},
  publisher={American Association for the Advancement of Science}
}

@article{blanchet2019robust,
  title={Robust Wasserstein profile inference and applications to machine learning},
  author={Blanchet, Jose and Kang, Yanjun and Murthy, Karthyek},
  journal={Journal of Applied Probability},
  volume={56},
  pages={830--857},
  year={2019}
}

@article{blanchet2025distributionally,
  title={Distributionally robust optimization and robust statistics},
  author={Blanchet, Jose and Li, Jiajin and Lin, Sirui and Zhang, Xuhui},
  journal={Statistical Science},
  volume={40},
  number={3},
  pages={351--377},
  year={2025},
  publisher={Institute of Mathematical Statistics}
}

@article{Blondel2008,
  title={Fast unfolding of communities in large networks},
  author={Blondel, Vincent D and Guillaume, Jean-Loup and Lambiotte, Renaud and Lefebvre, Etienne},
  journal={Journal of Statistical Mechanics: Theory and Experiment},
  volume={2008},
  number={10},
  pages={P10008},
  year={2008}
}

@book{boyd2004convex,
  title={Convex optimization},
  author={Boyd, Stephen and Vandenberghe, Lieven},
  year={2004},
  publisher={Cambridge University Press}
}

@article{breheny2011coordinate,
  title={Coordinate descent algorithms for nonconvex penalized regression, with applications to biological feature selection},
  author={Breheny, Patrick and Huang, Jian},
  journal={The Annals of Applied Statistics},
  volume={5},
  number={1},
  pages={232},
  year={2011}
}

@article{cai2011constrained,
  title={A constrained l1 minimization approach to sparse precision matrix estimation},
  author={Cai, Tony and Liu, Weidong and Luo, Xi},
  journal={Journal of the American Statistical Association},
  volume={106},
  number={494},
  pages={594--607},
  year={2011},
  publisher={Taylor \& Francis}
}

@article{candes2007dantzig,
    author = {Emmanuel Candes and Terence Tao},
    title = {The Dantzig selector: Statistical estimation when p is much larger than n},
    volume = {35},
    journal = {The Annals of Statistics},
    number = {6},
    publisher = {Institute of Mathematical Statistics},
    pages = {2313 -- 2351},
    year = {2007}
}

@article{chen2008extended,
  title={Extended Bayesian information criteria for model selection with large model spaces},
  author={Chen, Jiahua and Chen, Zehua},
  journal={Biometrika},
  volume={95},
  number={3},
  pages={759--771},
  year={2008},
  publisher={Oxford University Press}
}

@article{ChenRenZhaoZhou2016,
  title={Asymptotically normal and efficient estimation of covariate-adjusted Gaussian graphical model},
  author={Chen, Mengjie and Ren, Zhao and Zhao, Hongyu and Zhou, Harrison},
  journal={Journal of the American Statistical Association},
  volume={111},
  number={513},
  pages={394--406},
  year={2016},
  publisher={Taylor \& Francis}
}

@article{chicco2020advantages,
  title={The advantages of the Matthews correlation coefficient (MCC) over F1 score and accuracy in binary classification evaluation},
  author={Chicco, Davide and Jurman, Giuseppe},
  journal={BMC Genomics},
  volume={21},
  number={1},
  pages={6},
  year={2020},
  publisher={BioMed Central}
}

@inproceedings{cisneros2020distributionally,
  title={Distributionally robust formulation and model selection for the graphical lasso},
  author={Cisneros-Velarde, Pedro and Petersen, Alexander and Oh, Sang-Yun},
  booktitle={International Conference on Artificial Intelligence and Statistics},
  pages={756--765},
  year={2020},
  organization={PMLR}
}

@article{delage2010distributionally,
  title={Distributionally robust optimization under moment uncertainty with application to data-driven problems},
  author={Delage, Erick and Ye, Yinyu},
  journal={Operations Research},
  volume={58},
  number={3},
  pages={595--612},
  year={2010}
}

@article{duchi2021statistics,
  title={Statistics of robust optimization: A generalized empirical likelihood approach},
  author={Duchi, John C and Glynn, Peter W and Namkoong, Hongseok},
  journal={Mathematics of Operations Research},
  volume={46},
  number={3},
  pages={946--969},
  year={2021},
  publisher={INFORMS}
}

@article{erdos1959random,
  title={On random graphs I},
  author={Erd{\H{o}}s, Paul and R{\'e}nyi, Alfr{\'e}d},
  journal={Publicationes Mathematicae (Debrecen)},
  volume={6},
  pages={290--297},
  year={1959}
}

@article{fan2001variable,
  title={Variable selection via nonconcave penalized likelihood and its oracle properties},
  author={Fan, Jianqing and Li, Runze},
  journal={Journal of the American Statistical Association},
  volume={96},
  number={456},
  pages={1348--1360},
  year={2001},
  publisher={Taylor \& Francis}
}

@article{fan2016overview,
  title={An overview of the estimation of large covariance and precision matrices},
  author={Fan, Jianqing and Liao, Yuan and Liu, Han},
  journal={The Econometrics Journal},
  volume={19},
  number={1},
  pages={C1--C32},
  year={2016},
  publisher={Oxford University Press Oxford, UK}
}

@article{foygel2010extended,
  title={Extended Bayesian information criteria for Gaussian graphical models},
  author={Foygel, Rina and Drton, Mathias},
  journal={Advances in Neural Information Processing Systems},
  volume={23},
  year={2010}
}

@article{friedman2008sparse,
  title={Sparse inverse covariance estimation with the graphical lasso},
  author={Friedman, Jerome and Hastie, Trevor and Tibshirani, Robert},
  journal={Biostatistics},
  volume={9},
  number={3},
  pages={432--441},
  year={2008},
  publisher={Oxford University Press}
}

@article{gao2024wasserstein,
  title={Wasserstein distributionally robust optimization and variation regularization},
  author={Gao, Rui and Chen, Xi and Kleywegt, Anton J},
  journal={Operations Research},
  volume={72},
  number={3},
  pages={1177--1191},
  year={2024},
  publisher={Informs}
}

@article{goh2010distributionally,
  title={Distributionally robust optimization and its tractable approximations},
  author={Goh, Joel and Sim, Melvyn},
  journal={Operations Research},
  volume={58},
  number={4-part-1},
  pages={902--917},
  year={2010}
}

@article{Hirose2017,
  author  = {Hirose, Kenji and Fujisawa, Hironori and Sese, Jun},
  title   = {Robust sparse Gaussian graphical modeling},
  journal = {Journal of Multivariate Analysis},
  volume  = {161},
  pages   = {172--190},
  year    = {2017}
}

@article{hojsgaard2008graphical,
  title={Graphical Gaussian models with edge and vertex symmetries},
  author={H{\o}jsgaard, S{\o}ren and Lauritzen, Steffen L},
  journal={Journal of the Royal Statistical Society Series B: Statistical Methodology},
  volume={70},
  number={5},
  pages={1005--1027},
  year={2008},
  publisher={Oxford University Press}
}

@article{Huang2010,
    title = {Learning brain connectivity of Alzheimer's disease by sparse inverse covariance estimation},
    journal = {NeuroImage},
    volume = {50},
    number = {3},
    pages = {935-949},
    year = {2010},
    author = {Shuai Huang and Jing Li and Liang Sun and Jieping Ye and Adam Fleisher and Teresa Wu and Kewei Chen and Eric Reiman}
}

@inproceedings{johnson2012high,
  title={High-dimensional sparse inverse covariance estimation using greedy methods},
  author={Johnson, Christopher and Jalali, Ali and Ravikumar, Pradeep},
  booktitle={Artificial Intelligence and Statistics},
  pages={574--582},
  year={2012},
  organization={PMLR}
}

@article{kuhn2025distributionally,
  title={Distributionally robust optimization},
  author={Kuhn, Daniel and Shafiee, Soroosh and Wiesemann, Wolfram},
  journal={Acta Numerica},
  volume={34},
  pages={579--804},
  year={2025},
  publisher={Cambridge University Press}
}

@article{lam2009sparsistency,
  title={Sparsistency and rates of convergence in large covariance matrix estimation},
  author={Lam, Clifford and Fan, Jianqing},
  journal={The Annals of Statistics},
  volume={37},
  number={6B},
  pages={4254},
  year={2009}
}

@manual{covtools2025,
  title = {CovTools: Statistical Tools for Covariance Analysis},
  author = {Lee, Kyoungjae and Lin, Lizhen and You, Kisung},
  year = {2025},
  note = {R package version 0.5.6}
}

@article{li2021penalized,
  title={Penalized composite likelihood for colored graphical Gaussian models},
  author={Li, Qiong and Sun, Xiaoying and Wang, Nanwei and Gao, Xin},
  journal={Statistical Analysis and Data Mining: The ASA Data Science Journal},
  volume={14},
  pages={366--378},
  year={2021},
  publisher={Wiley Online Library}
}

@article{li2025Constraints,
author = {Li, Qiong and Wang, Nanwei and Gao, Xin and Pan, Jianxin},
title = {Bayesian Structure Learning for Graphical Models With Symmetry Constraints},
journal = {Biometrical Journal},
volume = {67},
number = {6},
pages = {e70091},
year = {2025}
}

@article{lingjaerde2024scalable,
  title={Scalable multiple network inference with the joint graphical horseshoe},
  author={Lingj{\ae}rde, Camilla and Fairfax, Benjamin P and Richardson, Sylvia and Ruffieux, H{\'e}l{\`e}ne},
  journal={The Annals of Applied Statistics},
  volume={18},
  number={3},
  pages={1899--1923},
  year={2024},
  publisher={Institute of Mathematical Statistics}
}

@article{liu2017tiger,
    author = {Han Liu and Lie Wang},
    title = {TIGER: A tuning-insensitive approach for optimally estimating Gaussian graphical models},
    volume = {11},
    journal = {Electronic Journal of Statistics},
    number = {1},
    publisher = {Institute of Mathematical Statistics and Bernoulli Society},
    pages = {241 -- 294},
    year = {2017}
}

@article{liu2009nonparanormal,
  title={The nonparanormal: Semiparametric estimation of high dimensional undirected graphs},
  author={Liu, Han and Lafferty, John and Wasserman, Larry},
  journal={Journal of Machine Learning Research},
  volume={10},
  pages={2295--2328},
  year={2009}
}

@article{liu2010stability,
  title={Stability approach to regularization selection (stars) for high dimensional graphical models},
  author={Liu, Han and Roeder, Kathryn and Wasserman, Larry},
  journal={Advances in Neural Information Processing Systems},
  volume={23},
  year={2010}
}

@article{liu2012high,
  title={High-dimensional semiparametric Gaussian copula graphical models},
  author={Liu, Han and Han, Fang and Yuan, Ming and Lafferty, John and Wasserman, Larry},
  journal={The Annals of Statistics},
  volume = {40},
  pages={2293--2326},
  year={2012},
  publisher={JSTOR}
}

@article{liu2015fast,
  title={Fast and adaptive sparse precision matrix estimation in high dimensions},
  author={Liu, Weidong and Luo, Xi},
  journal={Journal of Multivariate Analysis},
  volume={135},
  pages={153--162},
  year={2015},
  publisher={Elsevier}
}

@article{LohTan2018,
  author  = {Loh, Po-Ling and Tan, Xiaodong},
  title   = {High-dimensional robust precision matrix estimation: Cellwise corruption under $\epsilon$-contamination},
  journal = {Electronic Journal of Statistics},
  volume  = {12},
  number  = {1},
  pages   = {1429--1467},
  year    = {2018}
}

@article{lounici2011oracle,
author = {Karim Lounici and Massimiliano Pontil and Sara van de Geer and Alexandre B. Tsybakov},
title = {Oracle inequalities and optimal inference under group sparsity},
volume = {39},
journal = {The Annals of Statistics},
number = {4},
publisher = {Institute of Mathematical Statistics},
pages = {2164 -- 2204},
year = {2011}
}

@article{meinshausen2006high,
    author = {Nicolai Meinshausen and Peter B{\"u}hlmann},
    title = {High-dimensional graphs and variable selection with the Lasso},
    volume = {34},
    journal = {The Annals of Statistics},
    number = {3},
    publisher = {Institute of Mathematical Statistics},
    pages = {1436 -- 1462},
    year = {2006}
}

@article{esfahani2018data,
  title={Data-driven distributionally robust optimization using the Wasserstein metric: Performance guarantees and tractable reformulations},
  author={Mohajerin Esfahani, Peyman and Kuhn, Daniel},
  journal={Mathematical Programming},
  volume={171},
  number={1},
  pages={115--166},
  year={2018}
}

@misc{ds003720,
  author = {Tomoya Nakai AND Naoko Koide-Majima AND Shinji Nishimoto},
  title = {Music Genre fMRI Dataset},
  year = {2023},
  doi = {doi:10.18112/openneuro.ds003720.v1.0.1},
  publisher = {OpenNeuro}
}

@article{Newman2006,
  title={Modularity and community structure in networks},
  author={Newman, Mark EJ},
  journal={Proceedings of the National Academy of Sciences},
  volume={103},
  number={23},
  pages={8577--8582},
  year={2006},
  publisher={National Academy of Sciences}
}

@article{nguyen2022distributionally,
  title={Distributionally robust inverse covariance estimation: The Wasserstein shrinkage estimator},
  author={Nguyen, Viet Anh and Shafieezadeh-Abadeh, Soroosh and Kuhn, Daniel and Mohajerin Esfahani, Peyman},
  journal={Operations Research},
  volume={70},
  number = {1},
  pages = {490-515},
  year={2022},
  publisher={INFORMS}
}

@article{onizuka2025robust,
  title={Robust Bayesian graphical modeling using $\gamma$-divergence},
  author = {Takahiro Onizuka and Shintaro Hashimoto},
  journal={Journal of Multivariate Analysis},
  volume = {209},
  pages = {105461},
  year = {2025},
  publisher={Elsevier}
}

@article{panaretos2019statistical,
  title={Statistical aspects of Wasserstein distances},
  author={Panaretos, Victor M and Zemel, Yoav},
  journal={Annual Review of Statistics and Its Application},
  volume={6},
  pages={405--431},
  year={2019}
}

@article{Power2010,
  title={The development of human functional brain networks},
  author={Power, J.D. and Fair, D.A. and Schlaggar, B.L. and Petersen, S.E.},
  journal={Neuron},
  volume={67},
  number={5},
  pages={735--748},
  year={2010},
  publisher={Elsevier}
}

@article{rahimian2019distributionally,
  title={Distributionally robust optimization: A review},
  author={Rahimian, Hamed and Mehrotra, Sanjay},
  journal={arXiv preprint arXiv:1908.05659},
  year={2019}
}

@article{roverato2024exploration,
  title={Exploration of the search space of Gaussian graphical models for paired data},
  author={Roverato, Alberto and Nguyen, Dung Ngoc},
  journal={Journal of Machine Learning Research},
  volume={25},
  number={92},
  pages={1--41},
  year={2024}
}

@article{Ryali2012,
  title={Estimation of functional connectivity in fMRI data using stability selection-based sparse partial correlation with elastic net penalty},
  author={Ryali, Srikanth and Chen, Tianwen and Supekar, Kaustubh and Menon, Vinod},
  journal={NeuroImage},
  volume={59},
  number={4},
  pages={3852--3861},
  year={2012},
  publisher={Elsevier}
}

@article{sinha2018certifying,
  title={Certifying some distributional robustness with principled adversarial training},
  author={Sinha, Aman and Namkoong, Hongseok and Volpi, Riccardo and Duchi, John},
  journal={arXiv preprint arXiv:1710.10571},
  year={2017}
}

@article{sun2012scaled,
  title={Scaled sparse linear regression},
  author={Sun, Tingni and Zhang, Cun-Hui},
  journal={Biometrika},
  volume={99},
  number={4},
  pages={879--898},
  year={2012},
  publisher={Oxford University Press}
}

@article{sun2013sparse,
  title={Sparse matrix inversion with scaled lasso},
  author={Sun, Tingni and Zhang, Cun-Hui},
  journal={The Journal of Machine Learning Research},
  volume={14},
  number={1},
  pages={3385--3418},
  year={2013},
  publisher={JMLR. org}
}

@article{van2016distributionally,
  title={Distributionally robust control of constrained stochastic systems},
  author={Van Parys, Bart P and Goulart, Paul J and Morari, Manfred},
  journal={IEEE Transactions on Automatic Control},
  volume={61},
  number={2},
  pages={430--442},
  year={2016}
}

@article{wiesemann2014distributionally,
  title={Distributionally robust convex optimization},
  author={Wiesemann, Wolfram and Kuhn, Daniel and Sim, Melvyn},
  journal={Operations Research},
  volume={62},
  number={6},
  pages={1358--1376},
  year={2014}
}

@article{xue2012regularized,
  title={Regularized rank-based estimation of high-dimensional nonparanormal graphical models},
  author={Xue, Lingzhou and Zou, Hui},
  journal={The Annals of Statistics},
  volume={40},
  number={5},
  pages={2541--2571},
  year={2012},
  publisher={Institute of Mathematical Statistics}
}

@article{yuan2007model,
  title={Model selection and estimation in the Gaussian graphical model},
  author={Yuan, Ming and Lin, Yi},
  journal={Biometrika},
  volume={94},
  number={1},
  pages={19--35},
  year={2007},
  publisher={Oxford University Press}
}

@article{Zhang2019Aberrant,
  title={Aberrant brain connectivity in schizophrenia detected via a fast gaussian graphical model},
  author={Zhang, Aiying and Fang, Jian and Liang, Faming and Calhoun, Vince D and Wang, Yu-Ping},
  journal={IEEE Journal of Biomedical and Health Informatics},
  volume={23},
  number={4},
  pages={1479--1489},
  year={2018},
  publisher={IEEE}
}

@article{zhang2010nearly,
    author = {Cun-Hui Zhang},
    title = {Nearly unbiased variable selection under minimax concave penalty},
    volume = {38},
    journal = {The Annals of Statistics},
    number = {2},
    publisher = {Institute of Mathematical Statistics},
    pages = {894 -- 942},
    year = {2010}
}

@article{zhang2025multi,
  title={Multi-task learning for gaussian graphical regressions with high dimensional covariates},
  author={Zhang, Jingfei and Li, Yi},
  journal={Journal of Computational and Graphical Statistics},
  volume={34},
  number={3},
  pages={961--970},
  year={2025},
  publisher={Taylor \& Francis}
}

@article{zhao2012huge,
  title={The huge package for high-dimensional undirected graph estimation in R},
  author={Zhao, Tuo and Liu, Han and Roeder, Kathryn and Lafferty, John and Wasserman, Larry},
  journal={The Journal of Machine Learning Research},
  volume={13},
  number={1},
  pages={1059--1062},
  year={2012},
  publisher={JMLR. org}
}

@article{ZhengAllen2024,
  title={Graphical model inference with erosely measured data},
  author={Zheng, Lili and Allen, Genevera I},
  journal={Journal of the American Statistical Association},
  volume={119},
  number={547},
  pages={2282--2293},
  year={2024},
  publisher={Taylor \& Francis}
}

@article{zhu2009worst,
  title={Worst-case conditional value-at-risk with application to robust portfolio management},
  author={Zhu, Sanjay and Fukushima, Masao},
  journal={Operations Research},
  volume={57},
  number={5},
  pages={1155--1168},
  year={2009}
}

\end{document}